\documentclass[apj]{emulateapj}
\usepackage{amsmath}
\usepackage{natbib}

\begin{document}

\title{The Physical Conditions, Metallicity and Metal Abundance Ratios In a 
Highly Magnified Galaxy at \lowercase{z $=$ 3.6252} $^{*}$}

\author{Matthew B. Bayliss\altaffilmark{1,2},
Jane R. Rigby\altaffilmark{3}, 
Keren Sharon\altaffilmark{4},
Eva Wuyts\altaffilmark{5}, 
Michael Florian\altaffilmark{6,7},
Michael D. Gladders\altaffilmark{6,7}, 
Traci Johnson\altaffilmark{4} and
Masamune Oguri\altaffilmark{8,9} 
}

\newcommand{\arcname}{SGAS~J105039.6$+$001730}
\newcommand{\kms}{km s$^{-1}$}
\newcommand{\oiii}{O~{\small III}]}
\newcommand{\siiii}{Si~{\small III}]}
\newcommand{\ciii}{C~{\small III}]}
\newcommand{\logoh}{12 $+$ log({\small O/H})}

\newcommand{\firemag}{$27.3 ^{+11.4}_{-6.5}$}
\newcommand{\gmosmag}{$31.9 ^{+10.7}_{-7.8}$}
\newcommand{\photmag}{$31.4 ^{+10.8}_{-3.5}$}

\newcommand{\Px}{$-0.73 ^{+0.21}_{-0.14}$}
\newcommand{\Py}{$3.07 ^{+0.61}_{-1.32}$}
\newcommand{\Pe}{$0.10 ^{+0.04}_{-0.03}$}
\newcommand{\Ptheta}{$72.6 ^{+10.5}_{-2.9}$}
\newcommand{\Prc}{$81.2 ^{+12.0}_{-20.9}$}
\newcommand{\Psigma}{$1029 ^{+ 32}_{- 55}$}
\newcommand{\PsigmaA}{$492 ^{+ 43}_{- 48}$}
\newcommand{\PsigmaB}{$234 ^{+ 19}_{- 17}$}
\newcommand{\PsigmaC}{$124 ^{+ 18}_{- 10}$}
\newcommand{\PsigmaD}{$177 ^{+ 40}_{- 35}$}

\newcommand{\mstarnomag}{log(M$_{*}$) $=$ 11.0 $\pm$ 0.15 M$_{\odot}$ h$_{70}^{-1}$ }
\newcommand{\mstarmag}{log(M$_{*}$)  $=$ 9.5 $\pm$ 0.15 M$_{\odot}$ h$_{70}^{-1}$ }

\email{mbayliss@cfa.harvard.edu}

\altaffiltext{1}{Department of Physics, Harvard University, 17 Oxford St., 
Cambridge, MA 02138}
\altaffiltext{2}{Harvard-Smithsonian Center for Astrophysics, 60 Garden St., 
Cambridge, MA 02138}
\altaffiltext{3}{Observational Cosmology Lab, NASA Goddard Space Flight Center, 
Greenbelt, MD 20771}
\altaffiltext{4}{Department of Astronomy, the University of Michigan, 500 Church St. 
Ann Arbor, MI 48109}
\altaffiltext{5}{Max Plank Institute for Extraterrestrial Physics}
\altaffiltext{6}{Department of Astronomy \& Astrophysics, 
University of Chicago, 5640 South Ellis Avenue, Chicago, IL 60637}
\altaffiltext{7}{Kavli Institute for Cosmological Physics, 
University of Chicago, 5640 South Ellis Avenue, Chicago, IL 60637}
\altaffiltext{8}{Department of Physics, University of Tokyo, Tokyo 113-0033, Japan}
\altaffiltext{9}{Kavli Institute for the Physics and Mathematics of the Universe 
(Kavli IPMU, WPI), University of Tokyo, Chiba 277-8583, Japan}

\altaffiltext{*}{Based on observations from the Magellan Telescopes 
at Las Campanas Observatory, from Gemini Observatory, which is 
operated by the Association of Universities for Research in Astronomy, 
Inc., under a cooperative agreement with the NSF on behalf of the Gemini 
partnership: The United States, Canada, Chile, Australia, Brazil and 
Argentina, with additional supporting data obtained at the Subaru 
Telescope, which is operated by the National Astronomical Observatory of 
Japan, and on observations made with the NASA/ESA Hubble Space 
Telescope, obtained from the MAST data archive at the Space Telescope 
Science Institute, which is operated by the Association of Universities for 
Research in Astronomy, Inc., under NASA contract NAS 5-26555. 
These observations are associated with program \# GO13003.}

\begin{abstract}

We present optical and near-IR imaging and spectroscopy of \arcname, a strongly 
lensed galaxy at z $=$ 3.6252 magnified by $>$ 30$\times$, and derive its physical 
properties. We measure a stellar mass of log(M$_{*}$/M$_{\odot}$) 
$=$ 9.5 $\pm$ 0.35, star  formation rates from [O {\small II}]$\lambda$$\lambda$3727 
and H-$\beta$ of 55 $\pm$ 25 and 84 $\pm$ 24 M$_{\odot}$ yr$^{-1}$, respectively, 
an electron density of n$_{e} \leq$ 10$^{3}$ cm$^{-2}$, an electron temperature of 
T$_{e} \leq$ 14000 K, and a metallicity of \logoh~$=$ 8.3 $\pm$ 0.1. The strong C 
{\small III}]$\lambda$$\lambda$1907,1909 emission and abundance 
ratios of C, N, O and Si are consistent with well-studied starbursts at z $\sim$ 0 
with similar metallicities. Strong P Cygni lines and He~{\small II}$\lambda$1640 emission 
indicate a significant population of Wolf-Rayet stars, but synthetic spectra of individual 
populations of young, hot stars do not reproduce the observed integrated P Cygni 
absorption features. The rest-frame UV spectral features are indicative of a young 
starburst with high ionization, implying either 1) an ionization parameter 
significantly higher than suggested by rest-frame optical nebular lines, or 2) differences 
in one or both of the initial mass function and the properties of ionizing spectra of massive 
stars. We argue that the observed features are likely the result of a superposition of star 
forming regions with different physical properties. These results demonstrate the complexity 
of star formation on scales smaller than individual galaxies, and highlight the 
importance of systematic effects that result from smearing together the signatures of 
individual star forming regions within galaxies.

\end{abstract}


\keywords{gravitational lensing:strong --- galaxies: high-redshift 
galaxies --- galaxies: star formation}

\section{Introduction}

The accumulation of stellar mass and metallicity in galaxies at high 
redshift  involves complex interactions between several astrophysical 
processes, including star formation, winds and outflows, and gas accretion. 
In the current prevailing paradigm metal-poor gas is accreted from 
the intergalactic medium (IGM) and fuels star formation, which enriches 
the interstellar medium (ISM) with metals and generates galaxy-scale 
winds. This general picture provides a broad framework for understanding 
how galaxies build up their mass and metal content, but the details of how 
accretion, star formation, enrichment, and outflows are physically regulated 
remain poorly understood.

The rest-frame ultraviolet spectra of star forming galaxies include a wealth of 
diagnostics that constrain the properties of massive stars, the elemental abundances 
and physical properties of the nebular gas that those stars ionize, and the galaxy-scale 
outflows that they power. Rest-frame near-infrared spectra provide complementary 
measurements of the physical properties of the ionized nebular line-emitting regions. 
However, the observational data at z $>$ 3 remains extremely limited; the current 
challenge is to obtain good data for faint, distant galaxies.

\begin{deluxetable*}{ccccccc}
\tablecaption{Summary of Spectroscopic Observations\label{targettable}}
\tablewidth{0pt}
\tabletypesize{\tiny}
\tablehead{
\colhead{Telescope} &
\colhead{Instrument} &
\colhead{UT Date} &
\colhead{Disperser} &
\colhead{Filter} &
\colhead{Wavelengths (\AA)} &
\colhead{FWHM (\AA)} }
\startdata
Magellan-I & FIRE  & Jun 11,12 2011  & echelle   &  ---  &  10000-24000 & 2.8-6.1 \\
Gemini-North & GMOS  & Mar 29 2012  & R400 grating & OG515 & 5600-9800 & 8.2 \\
Magellan-I & IMACS  & Mar 17 2013  & 200 l/mm grism  & --- & 4800-9800  & 9.4  \\
Magellan-II & MagE  & May 06 2013  & echellette 175 l/mm  & --- &  3500-8500  & 1.2-2.0 \\
\enddata
\end{deluxetable*}

Photometry is relatively inexpensive, 
but reveals only limited constraints on the internal properties of galaxies. 
Spectroscopy provides significantly more information, but high S/N spectra of 
individual high redshift field galaxies are an expensive use of current 
instrumentation on larger-aperture telescopes \citep[e.g., ][]{erb2010}, and are 
limited to sampling the brightest (i.e., atypical) field galaxies. Stacking analyses 
of low signal-to-noise (S/N) spectra are useful for understanding the average 
properties of galaxies \citep[e.g.,][]{shapley2003,jones2012}, but sample neither the 
variations between galaxies nor the variations between regions within individual 
galaxies. Strong gravitational lensing provides a means by which the detailed physical 
properties of distant galaxies can be studied at high S/N. Galaxies that are highly 
magnified by massive foreground structures -- typically galaxy groups/clusters -- 
have their flux boosted by factors of $\gtrsim$ 10. Observations of these 
lensed sources with current facilities provide data that will only be matched, 
at best, by future generations of 30m class telescopes.


\begin{figure*}
\centering
\includegraphics[scale=0.445]{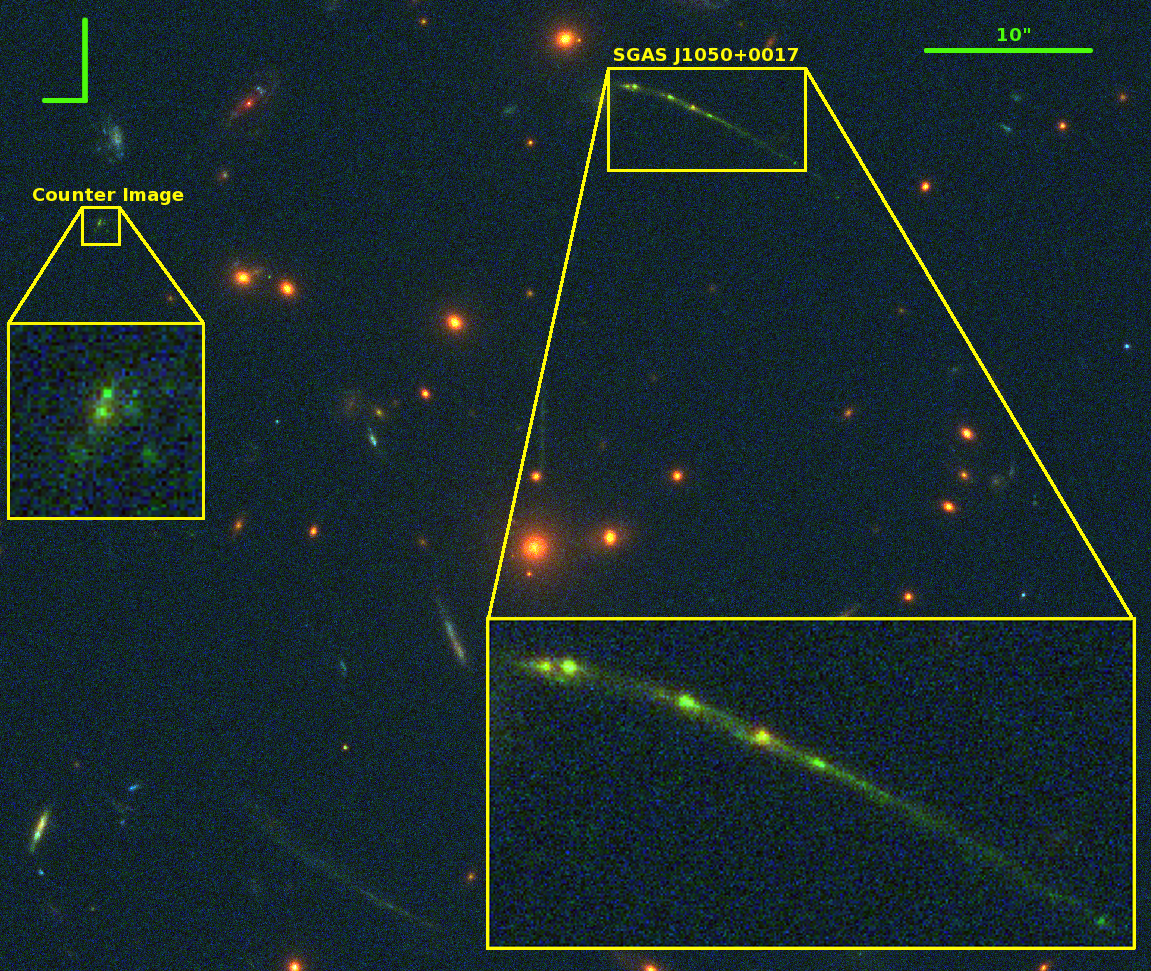}
\caption{\scriptsize{
Color composite image of the core of strong lensing cluster SDSS J1050+0017 
using {\it HST/WFC3} F110W/F606W/F390W assigned as R/G/B. We have labeled both 
\arcname~and its confirmed counter image, and provided zoom insets of the regions 
around both the arc and counter image. A compass in the upper left corner indicates 
North (long) and East (short), and the image scale is marked in the upper right corner.
}}
\label{fig:colorimage}
\end{figure*}

The exploitation of strong lensing to conduct high S/N studies of intrinsically faint 
galaxies at high redshift dates back to MS 1512-cB58 \citep[``cB58''][]{yee1996cb58}, 
a z $=$ 2.73 Lyman Break Galaxy (LBG) that is magnified by a foreground galaxy 
cluster by a factor of $\sim$ 30 \citep{seitz1998}. Follow-up observations of cB58 
yielded a high S/N spectra that revealed a tremendous level of detail about the 
chemical composition and state of the ISM \citep{pettini2000,teplitz2000,pettini2002}.
In recent years several ``cB58-like'' strongly lensed z $\sim$ 2-3 galaxies have 
been discovered \citep{fosbury2003,cabanac2005,allam2007,belokurov2007,
smail2007,diehl2009,lin2009,koester2010,wuyts2010}, and efforts to target these 
apparently bright (but intrinsically faint) galaxies for optical and NIR spectroscopy 
with large ground-based telescopes are accelerating. The lensing magnification allows 
for high quality spectroscopy from which the detailed physical properties and abundances 
can be derived \citep{hainline2009,quider2009,yuan2009,bian2010,dessauges2010,
erb2010,jones2010a,quider2010,dessauges2011,rigby2011,wuyts2012a,wuyts2012b,
jones2013a,stark2013a,shirazi2013,james2014}. 
Observations of emission line properties of fainter strongly lensed galaxies have 
also enabled measurements that push farther down the faint end of the 
luminosity function, deeper into the mass function, and out to higher redshifts 
\citep{bayliss2010,richard2011,christensen2012a,christensen2012b,wuyts2012a,
wuyts2012b,jones2013b}.

In this paper we present a multi wavelength analysis of \arcname, a LBG 
at z $=$ 3.6252 that is strongly lensed by a foreground galaxy cluster at z $=$ 0.593 
$\pm$ 0.002; the giant arc is highly magnified and has an apparent AB magnitude of 
F606W $=$ 21.48 (Figure~\ref{fig:colorimage}). This is the most distant 
galaxy for which a detailed study of the properties of massive stars and the 
inter-stellar medium (ISM) has been performed. \arcname~was discovered 
as a part of the Sloan Giant Arcs Survey (SGAS; M.~D.~Gladders et al. in prep), an 
on-going search for galaxy group and cluster scale strong lenses in the Sloan Digital 
Sky Survey \citep[SDSS;][]{york2000}. The full SGAS sample includes hundreds of strong 
lenses, many of which have been published 
in a variety of cosmological and astrophysical analyses \citep{oguri2009a,koester2010,
bayliss2010,bayliss2011a,bayliss2011b,gralla2011,oguri2012,bayliss2012,dahle2013,
wuyts2012a,wuyts2012b,gladders2013,blanchard2013}. \arcname~appears near the 
core of a strong lensing cluster that was first published by \citet{oguri2012}.

This paper is organized as follows: in $\S$~\ref{sec:data} we summarize the 
follow-up observations of \arcname~that inform this work, including both space- 
and ground-based imaging, as well as extensive optical and NIR spectroscopy. 
In $\S$~\ref{sec:methods} we review the analysis methods that we apply 
to the available data: spectral line profile measurements, 
systemic redshifts, strong lens modeling of the lens-source system, and 
photometry. $\S$~\ref{sec:analysis} includes the derivation of the detailed 
physical properties of \arcname, including stellar mass constraints from 
spectral energy distribution (SED) fitting, as well as the internal redding/extinction, 
star formation rates, electron density \& temperature, metallicity, and abundance 
ratios. In $\S$~\ref{sec:discussion} we discuss the physical picture that emerges 
for \arcname~given the available data, and we place it in the context of other 
studies of high redshift galaxies. Finally, $\S$~\ref{sec:conclusions} contains 
a summary of our analysis and its implications.

All magnitudes presented in this paper are in the AB system, 
based on calibration against the SDSS. We use a solar oxygen abundance 
Z$_{\odot}$: \logoh~$=$ 8.69 \citep{asplund2009}. All cosmologically dependent 
calculations are made using a standard flat $\Lambda$ cold dark matter 
($\Lambda$CDM) cosmology with $H_{0} = 70$ km s$^{-1}$ Mpc$^{-1}$, and 
matter density $\Omega_{M} = 0.27$ as preferred by observations of the 
Cosmic Microwave Background, supernovae distance measurements, 
and large scale structure constraints \citep{komatsu2011,reichardt2013}. 

\begin{figure}
\centering
\includegraphics[scale=0.37]{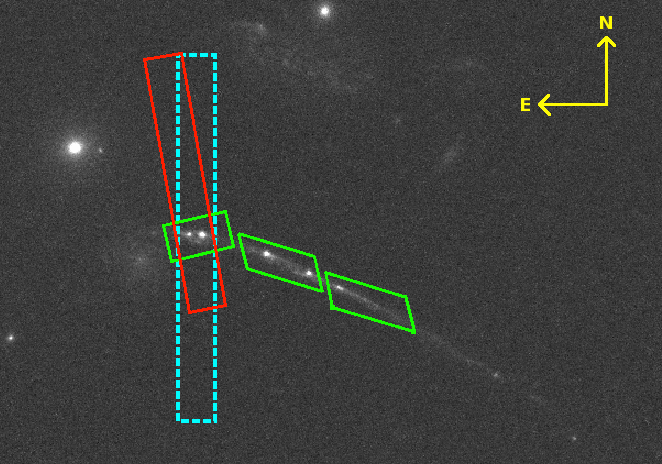}
\caption{\scriptsize{
{\it HST/WFC3-IR} F606W image of \arcname~with slits over-plotted for each of 
the follow-up spectroscopic observations described in Section~\ref{sec:data}. The 
three smaller green slits indicate the positions of slits placed on the arc in the 
GMOS nod-and-shuffle mask. The shorter GMOS slit was observed at both 
the pointing and nod positions of the N\&S observation, and therefore received 
twice the integration time as the other two tilted slits (this slit also covers the brightest 
part of the arc, and therefore dominates the signal in the final stacked GMOS spectrum). 
The longest cyan slit (dashed 
lines) indicates the position of the MagE slit, and the single long, tilted red slit 
indicates one of the two AB nod positions of the FIRE slit.}}
\label{slits_figure}
\end{figure}

\begin{figure*}[t]
\centering
\includegraphics[scale=0.52]{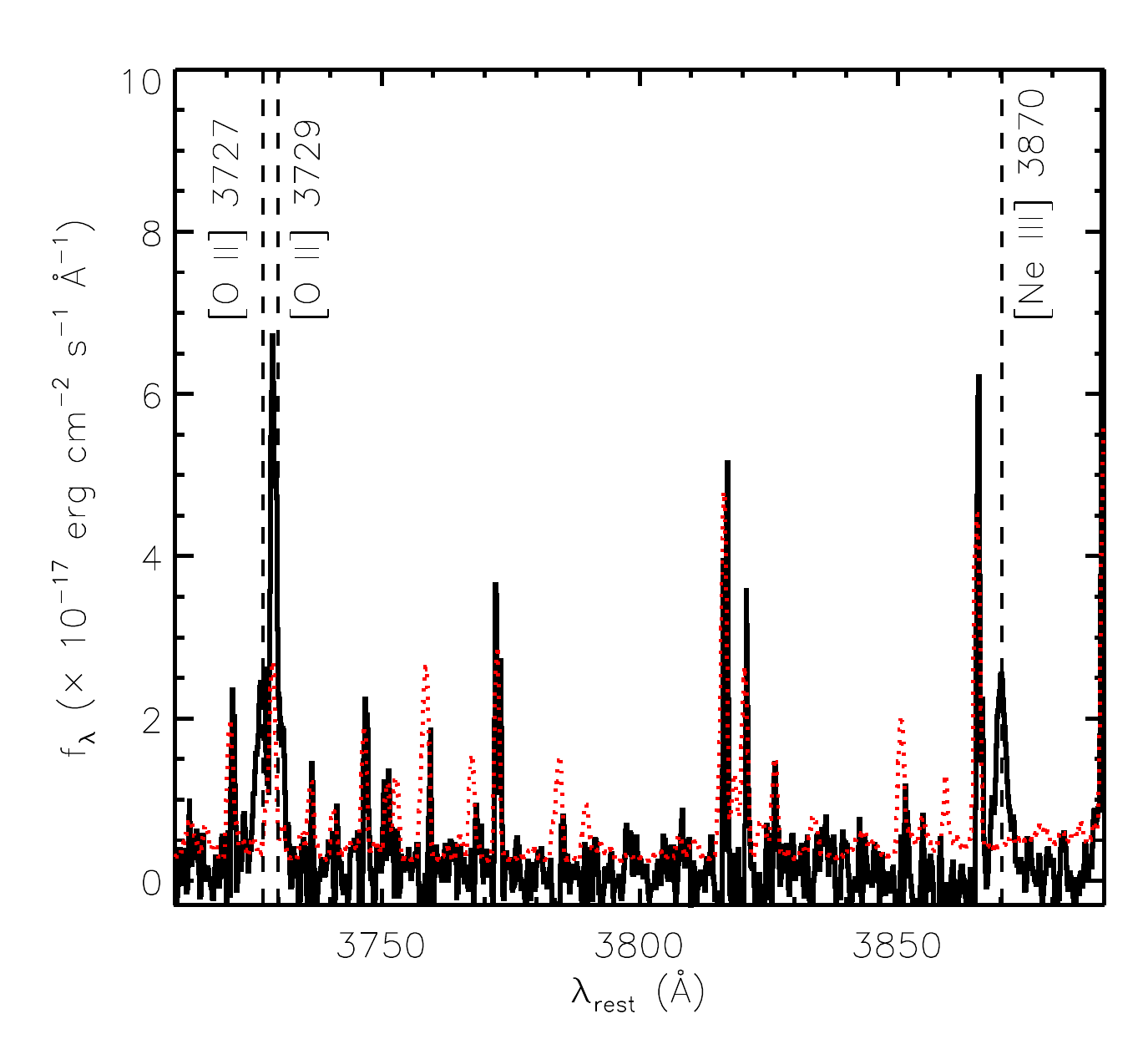}
\includegraphics[scale=0.52]{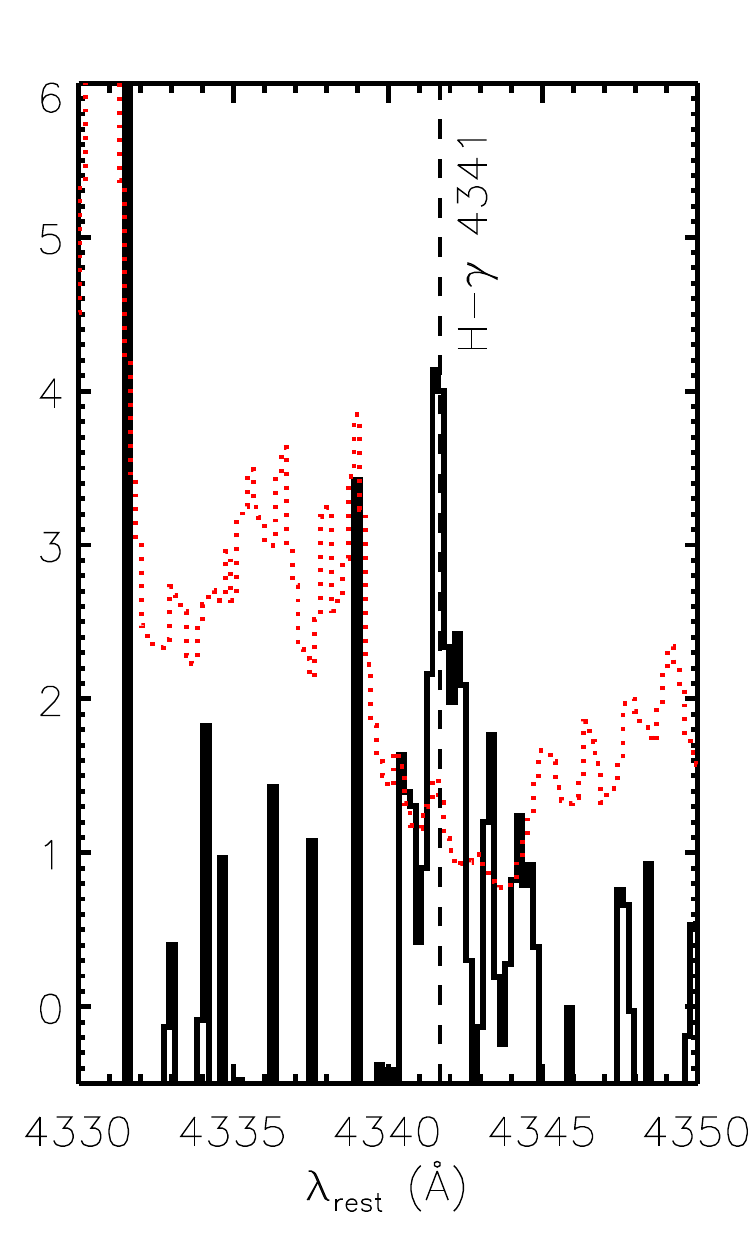}
\includegraphics[scale=0.52]{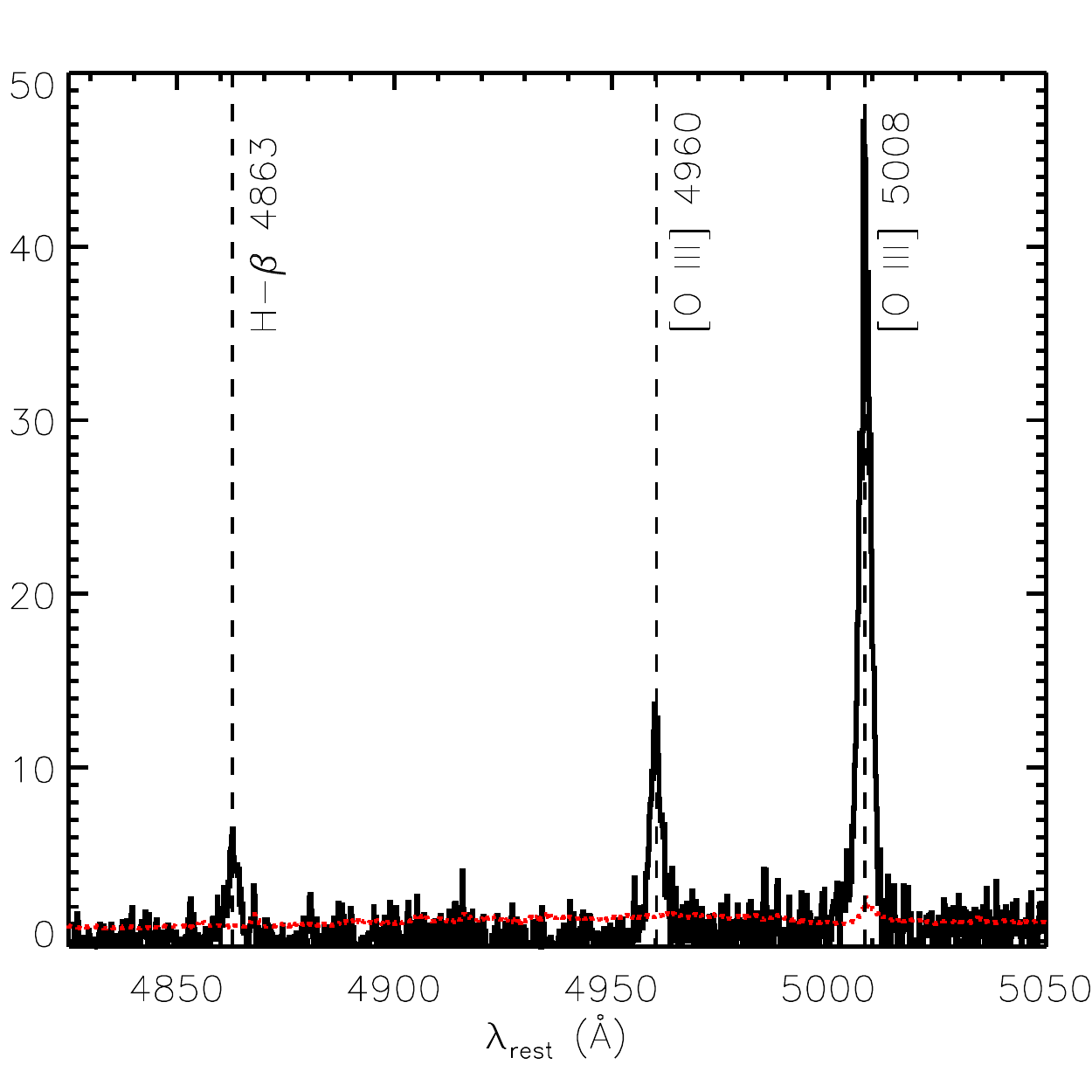}
\caption{\scriptsize{
Selected regions of the FIRE spectrum with emission lines indicated 
by vertical dashed lines and labeled, and the error array is over plotted as the dotted 
red line.
{\it Left:} FIRE spectrum covering the rest-frame wavelength range $\Delta \lambda =$ 
3710-3890-\AA.
{\it Middle:} FIRE spectrum covering the rest-frame wavelength range $\Delta \lambda =$ 
4330-4350\AA.
{\it Right:} FIRE spectrum covering the rest-frame wavelength range $\Delta \lambda =$ 
4825-5050\AA.
}}
\label{fig:firespec}
\end{figure*}

\section{Observations}
\label{sec:data}

\subsection{Imaging}
\label{sec:imaging}

\subsubsection{Subaru/SuprimeCam}

The field centered on \arcname~was observed with the Subaru telescope and the 
SuprimeCam instrument on UT Apr 7, 2011. The resulting data consist of $g$-, $r$- and 
$i$-band imaging, with total integration times of 1200~s, 2100~s, and 1680~s in $g$, $r$, and 
$i$, respectively. These observations were optimized to take advantage of the large field of 
view of SuprimeCam ($\sim$34\arcmin$\times$27\arcmin), and were used in a combined 
strong$+$weak lensing analysis of the foreground lensing cluster, SDSS J1050+0017; 
for details of the image reductions and resulting lensing analysis we direct the reader to  
\citet{oguri2012}.

\subsubsection{{\it HST/WFC3}}

\arcname~was observed with the Hubble Space Telescope and the Wide Field Camera 3 
using both the IR and UVIS channels on UT Apr 19-20, 2013 as a part of the GO 13003 
(PI: Gladders) program. Total integration times are 1212~s, 1112~s, 2400~s, 2388~s in the 
F160W, F110W, F606W and F390W filters, respectively. A post flash was applied to the 
individual F390W exposures to reduce the impact of charge-transfer inefficiency in the 
\emph{WFC3 UVIS} camera.

The image processing was performed using the 
\texttt{DrizzlePac}\footnote{http://drizzlepac.stsci.edu}  software package. Images 
were rescaled by re-drizzling the corrected flats with the \emph{astrodrizzle} routine 
to a scale of 0.03\arcsec\ pixel$^{-1}$ to take advantage of the finer grid made possible 
by our dithering pattern.  The resulting images were then aligned across filters with the 
image taken in the F606W filter using the \emph{tweakreg} function. The astrometric solutions 
provided by tweakreg were then propagated back to the corrected flat-fielded images using  
\emph{tweakback}.  Using \emph{astrodrizzle}, the F606W flat field images were drizzled onto 
a new grid with a scale of 0.03\arcsec\ pixel$^{-1}$, once with a drop size of 0.8 
pixels, and separately again with a drop size of 0.5. We found that a drop size of 0.5 for 
the IR camera and 0.8 for the UVIS camera provide the best sampling of the point spread 
function into our common pixel scale of $0\farcs03$.  
The resulting image was used as a reference grid for the re-drizzling of the images 
taken in the three remaining filters using the same scale and drop size and the 
updated astrometric solutions.

The WFC3 images were further processed to correct for IR `Blobs' not removed by 
the standard WFC3 pipeline flat-fields. These artifacts 
\footnote{http://www.stsci.edu/hst/wfc3/documents/ISRs/WFC3-2010-06.pdf} appear 
as small regions of reduced sensitivity due to dust particles contaminating the steering 
mirror that directs light into the WFC3 IR channel. We used object-masked individual 
frames in each IR filter, from the entire large HST GO program, to generate a sky flat 
for each filter. Though the IR blob artifacts are apparent in each flat, the raw flats are 
significantly contaminated from residual flux from real objects, and so we used 
GALFIT \cite{peng2010} to create a model of each IR blob as the sum of a few Gaussian 
components. This model is then used to flat-field the artifacts on each individual IR 
frame. UVIS flats were corrected for charge transfer inefficiencies using the CTE 
correction tool provided by STScI\footnote{www.stsci.edu/hst/wfc3/tools/cte\_tools}.

\subsubsection{{\it Spitzer/IRAC}}

Observations of \arcname~were obtained in the  3.6$\mu$m and 4.5$\mu$m 
channels as a part of {\it Spitzer} program \#70154 (PI: Gladders). Total 
integration times were 1200~s, taken during the warm Spitzer/IRAC mission. 
The data were reduced with the MOPEX software distributed by the Spitzer 
Science Center and drizzled to a pixel scale of 0.6\arcsec\ pixel$^{-1}$.

\subsection{Spectroscopy}
\label{sec:spectroscopy}

A summary of spectroscopic observations used in this work is shown in 
Table~\ref{targettable}. In the following sections we describe the spectroscopic 
data and their reduction.

\subsubsection{Magellan/FIRE}

Near-infrared spectroscopic observations were obtained with the 
Folded-port InfraRed Echellete (FIRE) instrument \citep{simcoe2013} on the 
Magellan-I (Baade) telescope. \arcname~
was observed on two nights, UT 2011-06-11 and 2011-06-12.  A pair of 
frames with individual integrations times of 602~s  were obtained each of 
the two nights, for a total integration of 2410~s. The echelle grating and the 
1.0~\arcsec\ slit were used, resulting in spectral resolution of R$=$3600 
(83 \kms). The position of the FIRE slit is shown in Figure~\ref{slits_figure}.

The A0~V star HD~96781 was observed as a standard star for purposes of 
fluxing and telluric correction. The star was observed immediately before 
each pair of science frames; the star was located 14 degrees from the science 
target, with sec(Z) airmass that differed by less than 0.2.

FIRE data were reduced using the FIREHOSE data reduction pipeline 
tools\footnote{wikis.mit.edu/confluence/display/FIRE/FIRE+Data+Reduction}, 
which were written in IDL by R.~Simcoe, J.~Bochanski, and M.~Matejek, and 
kindly provided to FIRE users by the FIRE team.

The FIRE pipeline uses lamp flat-fields to correct the pixel-to-pixel 
variation and sky flats to correct the illumination. The wavelength solution is fit using 
OH sky lines, and a two dimensional model of the sky is iteratively fit and subtracted 
following \citet{kelson2003}. The object spectrum is extracted using a spatial profile 
fit to the brightest emission line, and a flux calibration and corrected for telluric 
absorption correction are applied using the method of \citet{vacca2003} as 
implemented in an adapted version of the xtellcor routine from the SpeX pipeline. 
The final FIRE spectrum is a weighted average of the spectra that were extracted 
from the four individual exposures.

\begin{figure*}[t]
\centering
\includegraphics[scale=0.6]{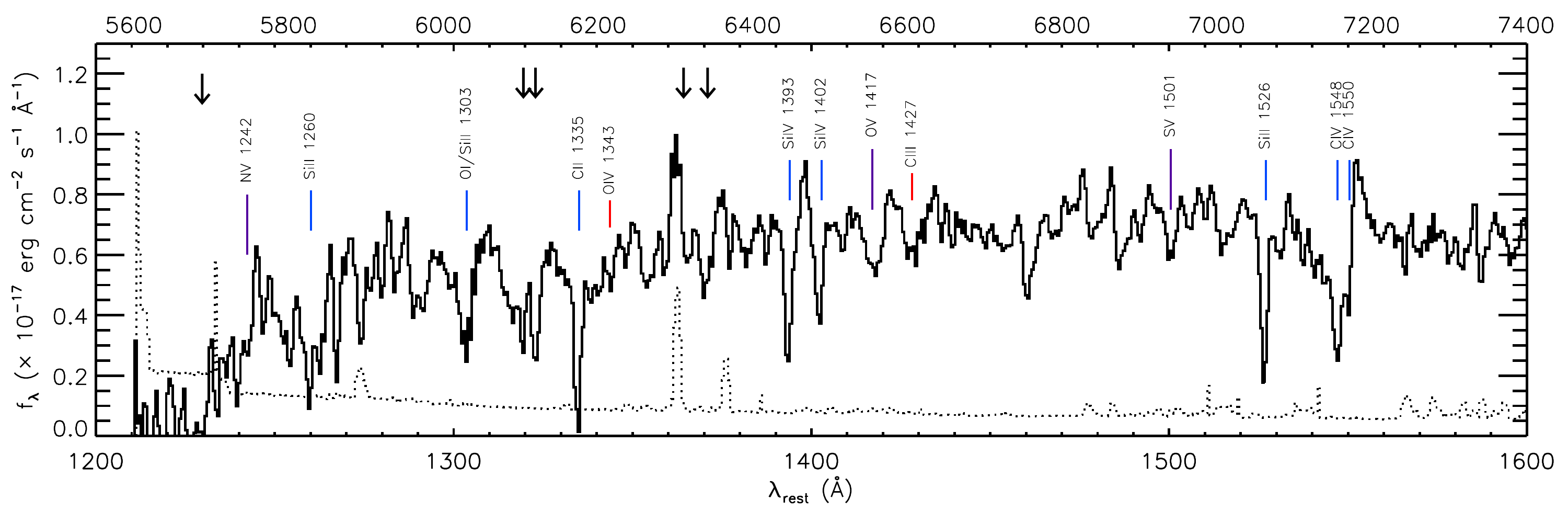}
\includegraphics[scale=0.6]{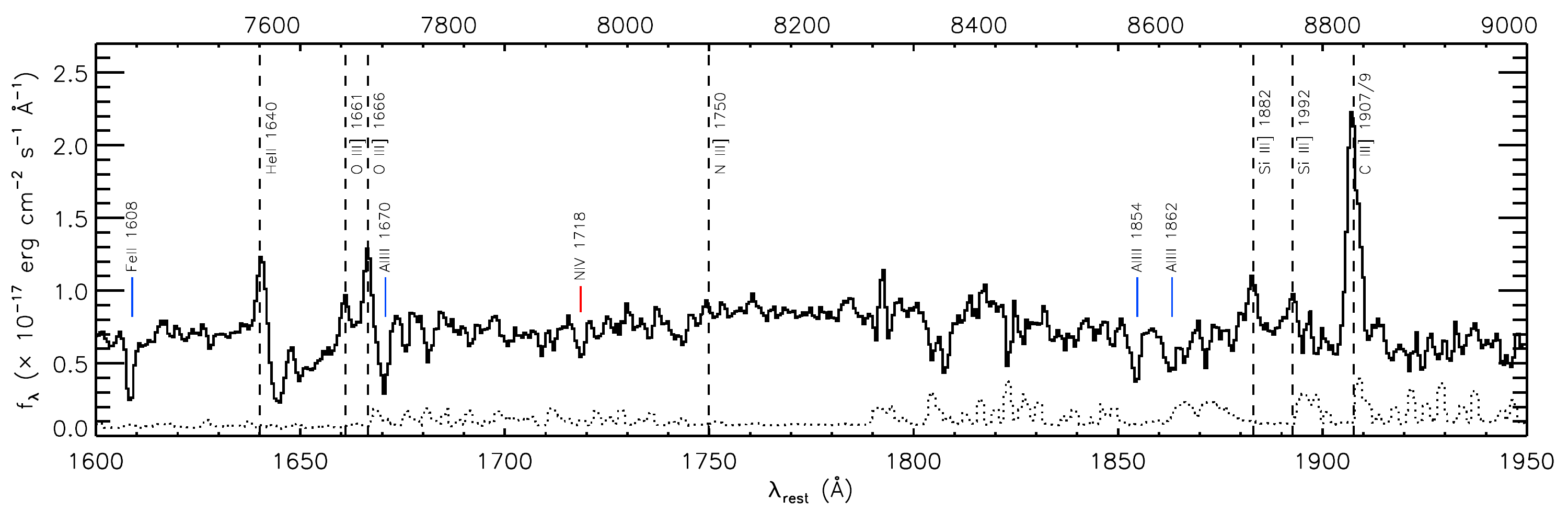}
\caption{\scriptsize{
{\it Top:} GMOS spectrum covering the rest-frame wavelength range $\Delta \lambda =$ 
1200-1600\AA. Spectral lines are indicated by type: black dashed lines are nebular 
emission lines, short solid red lines are stellar photospheric absorption features and 
medium length blue lines are ISM absorption lines, and long purple lines indicate 
transitions that could be either stellar photospheric or ISM (or more likely, a blend of 
the two). The error array is over plotted as the black dotted line, and the fit to the 
continuum level across the spectrum is plotted as a thin green line. The apparent emission 
feature that we observe at $\sim$6290\AA~is the result of a pernicious sky subtraction 
residual, and lines resulting from intervening absorption systems are indicated with 
downward facing arrows.
{\it Bottom:} GMOS spectrum covering the rest-frame wavelength range $\Delta \lambda =$ 
1600-1950\AA. Lines are indicated according to the same scheme as the top panel. The 
N III] 1750 emission line is only detected at  $\sim$2$\sigma$, but we indicate its location 
here because it is used later to constrain the relative nitrogen abundance. }}
\label{fig:gmosspec}
\end{figure*}

\subsubsection{Gemini/GMOS-N Nod-and-Shuffle}

\arcname~was observed with the Gemini-North telescope and the Gemini Multi-Object 
Spectrograph \citep[GMOS;][]{GMOS} on UT Mar 29 2012 in macro nod-and-shuffle (N\&S) 
mode and in clear conditions with seeing $\leq$0.75\arcsec\ as a part of queue program 
GN-2011A-Q-19. We used the R400\_G5305 grating in first order with the G515\_G0306 
long pass filter, and the detector binned by 2 in the direction of the spectral dispersion 
and unbinned spatially. The N\&S cycle length was 120s, chosen to reduce the number 
of shuffles in a given integration to mitigate charge trap effects. These observations were 
carried out after November 2011, and therefore used the new e2vDD detectors on 
GMOS-North, which provide significantly improved quantum efficiency relative to the 
older detectors that they replaced. 

A single multi-object slit mask was used to primarily target strongly lensed background 
sources in the core of SDSS J1050$+$0017; the approach is identical to the data described 
in detail by \citet{bayliss2011b} and we refer the reader to that paper for an in-depth 
description of the mask design. Four micro-slits on the mask were placed on \arcname, 
two of which covered the brightest knot in the arc at both the pointing and nod positions 
of the N\&S observations, and two slits which covered the length of the arc at only the 
pointing position (see Figure~\ref{slits_figure}). The total spectroscopic integration 
consisted of two 2400~s exposures, 
half of which was spent at each of the pointing and nod positions. The resulting spectra 
include 4800~s of total integration time on the brightest knot of \arcname, and 2400~s of 
total integration time on the fainter parts of the arc extending to the southwest. All science 
slits were 1\arcsec~wide, with the two slits extending along the length of the arc titled 30 
degrees relative to the dispersion axis to capture as much flux as possible. Slit placements 
for all of our spectroscopic observations are also shown in Figure~\ref{slits_figure}. The 
resolution of the data varies from R $\simeq$ 700-1100 (270-430 \kms).

Sky subtraction of N\&S data is simply a matter of differencing the two shuffled sections of 
the detector. The GMOS data were then wavelength calibrated, extracted, stacked, flux 
normalized, and analyzed using a custom pipeline that was developed using the 
XIDL\footnote{http://www.ucolick.org/$\sim$xavier/IDL/index.html} package. Flux 
calibration was performed using an archival observation of a single standard star 
and is therefore subject to pedestal offsets in the absolute flux calibration. The pipeline 
is almost identical to what was used by \citet{bayliss2011b}, with some updates made to 
account for the new e2vDD detectors.

\subsubsection{Magellan/IMACS}

We also observed the field centered on \arcname~with Magellan-I and the 
Inamori-Magellan Areal Camera \& Spectrograph (IMACS) using the long (f/2) camera 
on UT Mar 17 2013 in stable conditions with seeing ranging from $\sim$0.8-1.0\arcsec, 
with the airmass ranging from 1.26 to 1.15. Two multi-slit masks were each observed for 
3$\times$2400~s, and included several slits placed on potential 
counter-images of \arcname, as well as on other faint candidate strongly lensed 
background objects that were not targeted by the GMOS N\&S observations. 
We used the 200 l/mm grism and the spectroscopic (i.e., no order blocking) filter 
to allow for the broadest possible wavelength coverage and sensitivity. The 
detector was unbinned, resulting in spectral resolution R $\simeq$ 500-1000 
(300-460 \kms) and sensitivity over the wavelength range 
$\Delta\lambda =$ 4800-9800\AA. The data were wavelength calibrated, 
bias subtracted, flat-fielded, and sky subtracted using the COSMOS reduction 
package\footnote{http://code.obs.carnegiescience.edu/cosmos} designed specifically 
for IMACS and provided by Carnegie Observatories. The data were then extracted 
and stacked using custom IDL code; because the goal of these observations were 
redshift measurements, the extracted spectra were not precisely flux calibrated.

\subsubsection{Magellan/MagE}

We observed \arcname\ with the Magellan II (Clay) telescope and the Magellan Echellette 
(MagE) spectrograph \citep{marshall2008}. The observation started at May 6 2013 02:30:28 
UT, and the integration time was 3600s; the MagE slit was placed on the brightest knot of the 
arc, which had also previously been targeted with FIRE and GMOS (Figure~\ref{slits_figure}). 
The weather was clear, and the seeing 
as measured by wavefront sensing during the integration varied from 0.6 to 1.1\arcsec. 
The airmass rose from 1.30 to 1.57 during the observation. The target was acquired by 
blind-offsetting from a nearby brighter object; target acquisition was verified via the 
slit-viewing guider camera.  The slit was 1\arcsec\ by 10\arcsec~and the spectra were 
collected with 1$\times$1 binning resulting in a resolution of R $=$ 4100 (70~\kms).

The spectra were reduced using the LCO Mage pipeline written by D.~Kelson.  
The pipeline produces an extracted, one-dimensional, wavelength-calibrated spectrum 
for each echelle order.  The sensitivity function was computed using the IRAF 
tools onedspec.standard and sensfunc, using at least two observations each of 
the standard stars LTT 3864, EG~274, and Feige~67, at airmasses ranging from 1.02 
to 1.53. We scaled the sensitivity functions to the star with the highest 
throughput to create a composite sensitivity function.  To flux calibrate, these 
sensitivity functions were applied to the spectrum using the IRAF tool 
onedspec.calibrate. The uncertainty in the flux calibration is $\pm$20\%, 
resulting primarily from uncertainties in the slit losses due to variable seeing 
over the course of long spectroscopic integrations.
Overlapping orders of the echelle were combined with a weighted 
average to make a continuous spectrum, and then corrected to vacuum barycentric 
wavelength.  
The signal-to-noise ratio, per-pixel, of the continuum is low (rising from 0.1 at 
3500~\AA\ to 1 at 7500\AA.)

\subsection{GMOS vs MagE Flux Calibration}

The MagE spectrum was flux-calibrated more carefully, with several standard stars 
during the night, whereas the GMOS flux calibration was based on a 
non-contemporaneous (archival) standard star observation. We therefore extract  
the GMOS spectrum corresponding only from the slit which covers approximately the 
same region that was targeted by both the MagE and FIRE observations (see 
Figure~\ref{slits_figure}), and compare the resulting fluxed spectra against the MagE 
flux spectrum. In the wavelength range where both spectra have reasonable S/N 
(i.e., $\Delta \lambda \sim$ 6000-7000\AA) the two datasets have the same continuum 
flux level to within $\sim$5\%. Having performed this ``boot-strap'' flux calibration we 
feel secure in using the GMOS spectrum to measure line fluxes with a calibration that 
is accurate to within the uncertainty in the more carefully quantified MagE flux 
calibration ($\pm$20\%).


\begin{deluxetable}{clrrcc}
\tablecaption{Individual Line Redshifts\label{ztable}}
\tablewidth{0pt}
\tabletypesize{\tiny}
\tablehead{
\colhead{Ion} &
\colhead{$\lambda_{rest}$\tablenotemark{a} } &
\colhead{$\lambda_{obs}$} &
\colhead{$z$} & 
\colhead{Type\tablenotemark{b} } &
\colhead{Instrument}  
 }
\startdata
N {\small V} & 1242.804 &  5745.68 & 3.62316 &  2/4 & GMOS   \\
N {\small V} & 1242.804 &  5746.56 & 3.62386 &  2/4 & MagE   \\
Si {\small II} & 1260.422 &  5826.02 & 3.62228 &  1 & GMOS  \\
O~{\small I}  & 1302.169 &  6020.50 & 3.62344 &  1 & MagE  \\
O~{\small I}/Si {\small II} & 1303.860 &  6027.73 & 3.62299 &  1 & GMOS  \\
Si {\small II} & 1304.370 &  6031.87 & 3.62436  &  1 & MagE  \\
C {\small II}\tablenotemark{c} & 1335.480 &  6172.87 & 3.62221 &  1 & GMOS   \\
C {\small II}\tablenotemark{c} & 1335.480 &  6172.93 & 3.62225 &  1 & MagE  \\
O {\small IV} & 1343.514 &  6215.04 & 3.62596 &  2  &  GMOS  \\
Si {\small IV} & 1393.755 &  6444.76 & 3.62403 &  4 & GMOS \\
Si {\small IV} & 1393.755 &  6444.82 & 3.62407 &  4 & MagE  \\
Si {\small IV} & 1402.770 &  6486.14 & 3.62381 &  4 & GMOS   \\
O {\small V}\tablenotemark{d}  & 1417.866 & 6554.22 & 3.62260 &  2/4 & GMOS  \\
C {\small III} & 1427.839 &  6605.42 & 3.62617 &  2 & GMOS \\
S {\small V} & 1501.763 &  6939.89 & 3.62116 &  2/4 & GMOS \\
Si {\small II} & 1526.707 &  7060.20 & 3.62447 &  1 & GMOS   \\
Si {\small II} & 1526.707 &  7061.13 & 3.62508 &  1 & MagE   \\
C {\small IV} & 1548.203 &  7154.86 & 3.62139 &  4 & GMOS   \\
C {\small IV} & 1548.203 &  7156.04 & 3.62216 &  4 & MagE    \\
C {\small IV} & 1550.777 &  7170.61 & 3.62388 &  4 & GMOS  \\
C {\small IV} & 1550.777 &  7169.76 & 3.62333 &  4 & MagE   \\
Fe {\small II} & 1608.451 &  7438.93 & 3.62491 &  1 & GMOS   \\
Fe {\small II} & 1608.451 &  7437.89 & 3.62426 &  1 & MagE   \\
He {\small II} & 1640.420 &  7585.72 & 3.62426 &  3 & GMOS  \\
He {\small II} & 1640.420 &  7586.21 & 3.62455 &  3 & MagE  \\
O {\small III}] & 1660.809 &  7682.51 &  3.62577  &  3 & GMOS  \\
O {\small III}] & 1660.809 &  7681.81  &  3.62534 &  3  & MagE  \\
O {\small III}] & 1666.150 &  7707.79 & 3.62611 &  3 & GMOS  \\
O {\small III}] & 1666.150 &  7706.46 & 3.62531 &  3 & MagE   \\
Al {\small II} & 1670.788 &  7726.20 & 3.62429 &  1 & GMOS  \\
N {\small IV} & 1718.550 &  7948.66 & 3.62521 &  2 & GMOS   \\
Al {\small III} & 1854.716 &  8576.66 & 3.62425 &  1 & GMOS  \\
Al {\small III} & 1862.790 &  8615.79 & 3.62521 & 1 & GMOS  \\
Si {\small III}] & 1882.468 &  8708.81 & 3.62627 &  3 & GMOS  \\
Si {\small III}] & 1892.030 &  8754.13 & 3.62685 &  3 & GMOS  \\
C {\small III}] & 1907.640 &  8820.57 & 3.62381 &  3 & GMOS  \\
O [{\small II}]\tablenotemark{e} &  3727.092 & 17237.46 & 3.62491 &  3 & FIRE  \\
O [{\small II}]\tablenotemark{e}   & 3729.875 & 17250.17 & 3.62487 &  3 & FIRE  \\
Ne [{\small III}] & 3870.160 & 17899.61 & 3.62503 &  3 & FIRE  \\
H-$\gamma$ & 4341.684 & 20081.28 & 3.62521 &  3 & FIRE  \\
H-$\beta$ & 4862.683 & 22491.54 & 3.62534 &  3 & FIRE  \\
O [{\small III}] & 4960.295 & 22941.88 & 3.62510 &  3 & FIRE  \\
O [{\small III}] & 5008.240 & 23164.54 & 3.62528 &  3 & FIRE   \\
Mg {\small II}  & 2796.352 & 6102.39 & 1.18227 & 5 & GMOS  \\
Mg {\small II}  & 2803.531 & 6118.39 & 1.18239 & 5 & GMOS  \\
Al {\small II}  & 1670.788 & 5687.60  &  2.40389 &  5 & GMOS  \\
Al {\small III} & 1854.716 & 6309.98  &  2.40212 &  5  & GMOS  \\
Al {\small III} & 1862.790 & 6341.14  &  2.40413 & 5  & GMOS  
\enddata
\tablenotetext{a}{~Rest wavelengths are in a vacuum, taken from 
{\tiny www.pa.uky.edu/$\sim$peter/atomic/}}
\tablenotetext{b}{~Line type flag: 1 $=$ ISM, 2 $=$ Stellar, 3 $=$ Nebular, 4 $=$ ISM/P Cygni absorption, and 5 $=$ Intervening Absorption.}
\tablenotetext{c}{~Blend of the C II$\lambda$1334 and C II*$\lambda$1335 lines.}
\tablenotetext{d}{~Possibly blended with S V$\lambda$1501.}
\tablenotetext{e}{~Residuals from a bright sky line appear to be causing a systematic shift in the 
redshifts measured for these lines, as discussed in $\S$~\ref{sec:redshifts}.}
\end{deluxetable}

\section{Analysis Methods}
\label{sec:methods}
 
\subsection{Line Profile Measurements}
\label{sec:lprofiles}

We fit gaussians to all spectroscopic features of interest; the fits use a single 
gaussian profile with three free parameters: the normalization, width and centroid. 
For the FIRE data this process is performed directly on the extracted 1-dimensional 
spectrum, in which any continuum emission is consistent with zero flux to within the 
uncertainties of the data. The regions in the FIRE spectrum in which emission lines 
appear are shown in Figure~\ref{fig:firespec}.

The GMOS spectrum for \arcname~exhibits strong features in both absorption and 
emission (Figure~\ref{fig:gmosspec}). We begin our analysis of this spectrum by 
fitting a polynomial continuum model to the continuum. The continuum fitting process 
begins by identifying regions of continuum emission with good S/N (e.g., $\lambda_{obs} 
\sim$ 6650-6750\AA, 7250-7400\AA, 7500-7550\AA, 7800-8200\AA, and 
8900-9000\AA) and then iteratively add new wavelength ranges to to the fit. 
Our final continuum fit is robust to the exact model parameterization, and is 
plotted on top of the GMOS data in Figure~\ref{fig:gmosspec}. The residuals of this 
continuum fit are consistent with the uncertainties in the extracted GMOS spectrum. 
The continuum fit is subtracted from the GMOS spectrum prior to the measurement of 
individual line positions and fluxes. We then fit gaussian profiles to the 
contintuum-subtracted spectrum to measure their wavelength centroids, as well as -- in 
the case of emission features -- their fluxes. We incorporate both statistical (measurement) 
and systemic uncertainties into the emission line flux measurements. We determined 
the systematic uncertainty contribution from the continuum fit empirically by comparing 
the line flux measurements that result from different continuum models. The different 
continuum fits agree well, and the magnitude of the systemic contribution to the total 
uncertainty is typically $\lesssim$20\% that of the measurement uncertainty.

The MagE spectrum also includes continuum 
emission, though at lower S/N than in the GMOS data. We use the same procedure 
to fit a continuum model to the MagE data before measuring line profiles. All line flux 
measurements in the optical are made using the GMOS data, which is much higher 
S/N. We do measure some line profiles in the MagE spectra, and find centroids that 
agree well with the GMOS measurements of the same features. All measured line 
centroids are reported in Table~\ref{ztable}.

For measurements and analyses that span different spectra (e.g., comparing line 
fluxes between the GMOS and FIRE data), we restrict the analysis to the GMOS 
spectrum extracted only from the slit targeting the same bright knot as the MagE and 
FIRE observations. The total integration time for this slit was twice as long as the other 
slits, and the knot in question is the brightest part of the arc, so that the GMOS data 
from this slit alone provide a spectrum with only marginally lower S/N than a stack 
of all the GMOS slits. Comparing spectral features measured from this knot minimizes 
the geometric corrections between the different spectral datasets. 

Geometric corrections do not account for the effects of differential refraction, but the 
differential refraction effects across the GMOS spectrum should be minimal given the 
wavelengths covered by the observations. Atmospheric dispersion effects are also not 
a significant problem in spectra in the NIR. 

\subsection{Systemic Redshift Measurements}
\label{sec:redshifts}

Given the broad wavelength coverage (rest frame UV to optical) and good S/N of the 
spectra, it is possible to measure systemic redshifts for emission and absorption 
features with different astrophysical origins within \arcname. Typical uncertainties in 
individual line redshifts for well-detected transitions are $\sigma_{z} =$ 0.0006, 0.0002, 
and 0.0002 in the GMOS, MagE, and FIRE spectra, respectively (some low S/N lines have 
larger uncertainties, e.g., $\sigma_{z} =$ 0.001-0.002). These values include the 
uncertainties in the individual line centroids, as well as the (negligible) rms uncertainty 
in the wavelength calibration.

The first line system that we examine is the family of nebular emission lines that appear 
in both the optical (rest-UV) and NIR (rest-optical) spectra. These lines originate from 
ionized regions within the galaxy, i.e. HII regions around massive stars. From 14 
measurements of 14 separate nebular emission features we measure a systemic 
redshift $z_{neb} =$ 3.6253 $\pm$ 0.0008 (\oiii$\lambda$$\lambda$1661,1666 and 
He{\small II}$\lambda$1640 are detected in both the GMOS and MagE data). We exclude 
the [O {\small II}]$\lambda$$\lambda$3727,3729 doublet lines from inclusion in this 
systemic redshift due to their 
coincidence in wavelength with a bright sky line. The sky subtraction residuals from the 
bright sky line seem to cause an over-subtraction on the blue side of the sky line and an 
under-subtraction on the red side, which seems to result in a redshift measurements for 
the [O {\small II}]$\lambda$$\lambda$3727,3729 lines that is slightly biased low 
(see also $\S$~\ref{sec:electrondensity}~below).

There are also numerous absorption lines in the optical spectra that originate from 
ionized metals in the ISM of \arcname. These include a significant 
P Cygni absorption/emission profile for the C{\small IV}$\lambda$$\lambda$1448,1450 
doublet, and similar P Cygni features are also apparent at lower significance for the 
Si{\small IV}$\lambda$$\lambda$1393,1402 doublet. From 20 measurements of 14 
individual line features we measure a systemic ISM absorption redshift $z_{ISM} =$ 
3.6236 $\pm$ 0.0011. We also compute the systemic redshift for the 7 detected P 
Cygni absorption features alone, and find $z_{P-Cyg} =$ 3.6232 $\pm$ 0.0011 -- 
this is marginally more blueshifted than the complete set of ISM lines, which would 
be consistent with the P Cygni features tracing regions with stronger outflows, though 
the offset is not statistically significant.

There are also several spectral features that can arise from a blend of stellar and nebular 
P Cygni features, including N{\small V}$\lambda$$\lambda$1238,1242 and O {\small V} 
1371, O {\small V} 1417, S {\small V}$\lambda$1501 -- where O {\small V} 1417 can 
also be blended with Si {\small IV} 1417. We lack the S/N and spectral resolution 
to disentangle these features in the GMOS and MagE spectra, so we do not use 
them to compute any systemic redshifts, and flag them as possibly being both stellar 
and P Cygni (i.e., likely a blend of the two) in origin in (see Table~\ref{ztable}). The emission 
parts of the P Cygni features are difficult to fit because of significant asymmetry due to 
the neighboring absorption. These features may also originate, in part, from nebular 
line emission from these transitions (see $\S$~\ref{sec:pcygni}). Given the difficulties 
we refrain from measuring line centroids for the P Cygni emission, as it is not clear 
how to interpret such measurements.

Additionally, in the GMOS spectrum we note the presence of O{\small IV}$\lambda$1343, 
C {\small III}$\lambda$1427, and N{\small IV}$\lambda$1718 in 
absorption. All of these lines are associated with photospheric absorption in the 
atmospheres of massive stars \citep[e.g.,][]{pettini2000}, and are therefore likely tracing 
the stellar content of \arcname. These four lines are all weak relative 
to the much stronger ISM absorption lines, and have a mean redshift 
$z_{stars} =$ 3.6258 $\pm$ 0.0005.

All of the three systemic redshift measurements agree within 2$\sigma$ given the 
measurement uncertainties, but we note that the ISM absorption line redshift is 
formally blueshifted by 100 $\pm$ 70 \kms. The nebular and stellar photospheric 
redshifts agree within 1$\sigma$, as is expected given that both of these features 
should trace structures that are gravitationally bound within the galaxy.

To obtain the best possible systemic redshift constraint for \arcname~we restrict ourselves 
to using nebular emission lines in the FIRE spectrum. The FIRE spectrum is high 
resolution and includes many high S/N lines, whereas the nebular lines in the GMOS 
and MagE data are much lower spectral resolution and S/N, respectively. Computing the 
systemic redshift using just the remaining five lines in the FIRE spectrum 
results in a statistically consistent redshift measurement but reduces the uncertainty 
signficantly. The systemic nebular emission line redshift derived from the FIRE data 
is $z_{sys} =$ 3.6252 $\pm$ 0.0001. 

The telluric A band absorption feature is prominent in the GMOS spectrum, but at the 
redshift of \arcname, it fortuitously falls between the He~II$\lambda$1640 and 
O III]$\lambda$1661 emission lines without significantly affecting the flux of either line.
Because we use archival standard star observations to flux calibrate the GMOS spectrum, 
we cannot apply a reliable telluric absorption correction and so instead we refrain from 
using the affected parts of the GMOS spectrum.

All individual line redshifts are all presented in Table~\ref{ztable}, and the resulting 
systemic redshift measurements are given in Table~\ref{zsystemic}. 

\subsection{Intervening Absorption Systems}

In addition to the lines that area associated with \arcname, we also identify several 
foreground intervening features. An absorption doublet at $\lambda \sim$ 
6105\AA~is MgII$\lambda$$\lambda$2796,2803 from an intervening galaxy at z $=$ 
1.1820 $\pm$ 0.0002. There is also an absorption doublet at $\lambda \sim$ 6330\AA~that we 
identify as the Al {\small III}$\lambda$$\lambda$1854,1862 doublet, along with a 
an absorption line at 5687\AA~that we identify as Al {\small II}$\lambda$1670; these 
Al lines originate from a second intervening galaxy at z $=$ 2.4034 $\pm$ 0.0011. 
Interestingly, both of these intervening galaxies are also strongly lensed by the foreground 
cluster, SDSS~J1050+0017 and have redshifts confirmed from spectroscopic data 
that are not presented in this paper. Several other absorption features that are present 
in the GMOS spectrum remain unidentified -- at least some of these likely result from 
intervening absorption by the foreground galaxy labeled ``Gal 1'' in 
Figure~\ref{fig:lensmodel} for which we do not yet have a spectroscopic redshift 
measurement. Lines from intervening absorbers are include in Table~\ref{ztable} and 
Table~\ref{zsystemic}.

\begin{deluxetable}{lccc}
\tablecaption{Systemic Redshifts\label{zsystemic}}
\tablewidth{0pt}
\tabletypesize{\tiny}
\tablehead{
\colhead{System} &
\colhead{$z$} & 
\colhead{ $\sigma_{z}$ } &
\colhead{ \# Lines}  
 }
\startdata
All Nebular  & 3.6253  & 0.0008 &  14  \\
IR Nebular\tablenotemark{a} & 3.6252  & 0.0001 &  5  \\
Stellar &  3.6258  & 0.0005 &  3  \\
ISM  &  3.6236  &  0.0011  &  20  \\
P Cygni Abs & 3.6232  & 0.0011  &  7 \\
\hline
Intervening &  &  &  \\
Absorbers\tablenotemark{b} &  &  &  \\
\hline
Arc A & 2.4034  &  0.0011  &  3  \\
Gal 3 Arc & 1.1823 & 0.0001  &  2  
\enddata
\tablenotetext{a}{~Redshift derived only using 5 nebular lines in the FIRE spectrum.}
\tablenotetext{b}{~Names track back to objects in Figure~\ref{fig:lensmodel}.}
\end{deluxetable}

\subsection{Spatially Extended Spectral Emission and Lack of AGN Features}

\begin{figure}[b]
\centering
\includegraphics[scale=0.54]{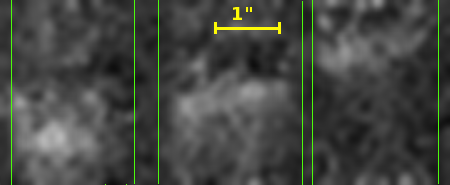}
\caption{\scriptsize{
The 2D GMOS spectrum of \arcname~is shown, zoomed in on the region where the 
C III] doublet appears and smoothed with a gaussian kernel with $\sigma = 1$ detector 
pixel. The vertical axis here is the direction of dispersion, the horizontal axis is 
the spatial direction along the detector, and the vertical green lines indicate the 
approximate edges of the three N\&S slitlets that are plotted in Figure~\ref{slits_figure} 
with the spatial scale is indicated by the yellow bar. The C III] emission feature appears 
in all three N\&S slitlets, and is clearly extended along the length of each slitlet.}}
\label{fig:gmos2D}
\end{figure}

From the GMOS observations of \arcname~we have optical spectra that extend along 
the length of the arc, which allows us to test for spatial variations in the spectrum. There 
are several emission lines that are strong enough to be well-detected in spectra that are 
extracted from sub-apertures along the GMOS slits covering the arc, and we can look 
for variations in the relative strengths of these lines as a function of spatial position. 
\ciii$\lambda$$\lambda$1907,1909 and He{\small II}$\lambda$1640 are the two 
highest S/N such lines, and we find no evidence for spatial variation in their equivalent 
widths in the GMOS data. The seeing during the GMOS observations was 0.65\arcsec, 
and we also note that the emission line features are clearly extended along the GMOS 
slits, which are between 1.5\arcsec~and 2\arcsec long (Figure~\ref{fig:gmos2D}). This 
spatially extended 
emission rules out and active galactic nucleus (AGN) as the dominant source of ionizing 
photons in \arcname. There is also a notable lack of emission lines that would be 
associated with the extremely hard ionizing spectrum of an AGN, such as N {\small V} in 
the rest-UV and [Ne {\small V}] in the rest-optical. We can also use the emission lines 
in the FIRE spectrum to evaluate where \arcname~falls in the log([O II] 3727/ [Ne III] 3869) 
vs log(O$_{2Ne3}$) diagnostic space from \citet{perezmontero2007}. With 
log([O II] 3727/ [Ne III] 3869) $= -0.4$ and log(O$_{2Ne3}$) $= 0.9$, \arcname resides 
well within the region occupied by star forming objects in this space. Based on all of 
the available information we therefore conclude that AGN activity is not powering a 
significant fraction of the ionized gas emission, and proceed with the assumption 
that AGN contributions to the spectrum of \arcname~are negligible. 

\subsection{PSF Matched Photometry}

Photometry of the giant arc is performed using custom IDL code that lets us construct 
apertures which follow the ridge-line of the arc. Images are convolved to a common 
point spread function (PSF) so that the apertures can be defined and applied to the 
same regions on the sky. This process is described in more detail in \citet{wuyts2010} 
and \citet{bayliss2012}. \arcname~has a total magnitude of F606W $=$ 21.48 $\pm$ 0.05 
and $i =$ 21.18 $\pm$ 0.06.

\subsection{Lensing Analysis}

\subsubsection{Strong Lens Model}
\label{sec:slmodel}

\begin{figure*}
\centering
\includegraphics[scale=0.75]{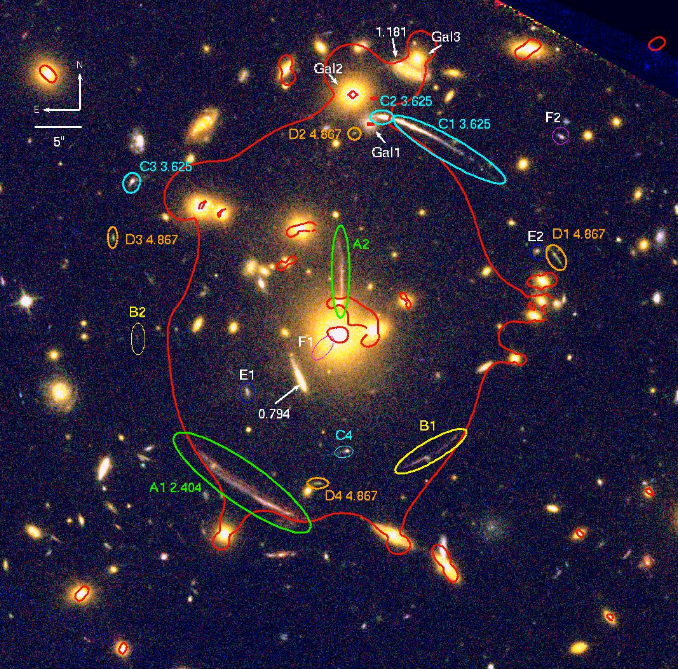}
\caption{\scriptsize{
Color composite image of the core of strong lensing cluster SDSS J1050+0017 
using {\it HST}/WFC3 F160W,F110W,F606W. The critical curve of the best-fit lens 
model is plotted in red. Multiply-lensed galaxies are marked with ellipses, and their 
IDs and redshifts are labeled. The thick ellipses mark confirmed arcs, and thin ellipses 
mark arc candidates that were predicted by the lens model but not spectroscopically 
confirmed. The foreground galaxies that were individually modeled as contributing to 
the strong lensing are indicated. We also indicate other objects in the background of 
the clusters for which we measured spectroscopic redshifts. See $\S$~\ref{sec:slmodel} 
for a complete description of the strong lens model.}}
\label{fig:lensmodel}
\end{figure*}

In addition to \arcname, we identify several different strongly lensed background 
galaxies around the core of the foreground galaxy cluster; these are labeled in 
Figure~\ref{fig:lensmodel}. Galaxy A, at z $=$ 2.404, is lensed into a giant tangential 
arc (A1) and a radial arc (A2), both with similar morphology and colors in the 
\emph{HST} and \emph{Spitzer} data. Galaxy B is a faint tangential arc at unknown 
redshift. Galaxy D is lensed into four images (D1-4), spectroscopically confirmed to 
be at z $=$ 4.867 from Gemini/GMOS (D1) and Magellan/IMACS (D2,3,4).

Two images of galaxy C form the giant 
arc \arcname. One counter image (C3) at ($\alpha$,$\delta$) = 10:50:41.336, 
$+$00:17:23.35 (J2000) was spectroscopically confirmed by IMACS based on the 
presence of emission line features coincident with He~{\small II}$\lambda$1640, 
C~{\small IV}$\lambda$$\lambda$1448,1450 and \ciii$\lambda$$\lambda$1907,1909 
at the same redshift and in the same approximate relative strengths as
observed in the GMOS spectrum of the main arc. Another image (C4) is 
predicted by the lens model and visually confirmed in the HST data, but 
lacks spectroscopic confirmation.

The positions and redshifts of the spectroscopically confirmed galaxies are 
used as constraints in the lens modeling process. With the exception of system 
C (\arcname), we use one position per image. For system C, we use the 
positions of the four brightest knots in C1 and C2, and the two brightest knots 
in C3.The lens model is computed using the publicly-available software, 
\texttt{Lenstool} \citep{jullo2007}, utilizing a Markov Chain Monte Carlo 
minimizer both in the source plane and the image plane. The lens is 
represented by several pseudo-isothermal ellipsoidal mass distribution 
(PIEMD) halos, described by the following parameters: position $x$, $y$; a 
fiducial velocity dispersion $\sigma$; a core radius r$_{core}$; a cut radius 
r$_{cut}$; ellipticity $e = (a^{2} - b^2{)}/(a^{2} + b^{2})$, where $a$ and $b$ are 
the semi major and semi minor axes, respectively; and a position angle 
$\theta$. The PIEMD profile is formally the same as dual Pseudo Isothermal 
Elliptical Mass Distribution \citep[dPIE, see][]{eliasdottir2007}. 

\begin{deluxetable*}{lccccccc} 
 \tablecolumns{8} 
\tablecaption{Best-fit lens model parameters  \label{table.lensmodel}} 
\tablehead{\colhead{Halo }   & 
            \colhead{RA}     & 
            \colhead{Dec}    & 
            \colhead{$e$}    & 
            \colhead{$\theta$}       & 
            \colhead{$r_{\rm core}$} &  
            \colhead{$r_{\rm cut}$}  &  
            \colhead{$\sigma_0$}\\ 
            \colhead{(PIEMD)}   & 
            \colhead{($\arcsec$)}     & 
            \colhead{($\arcsec$)}     & 
            \colhead{}    & 
            \colhead{(deg)}       & 
            \colhead{(kpc)} &  
            \colhead{(kpc)}  &  
            \colhead{(km s$^{-1}$)}  } 
\startdata 
Cluster    & \Px        & \Py           & \Pe       & \Ptheta        &\Prc     & [1500]    & \Psigma  \\ 
BCG         & [0.0]       & [0.0]        & [0.067] & [-81.3]          & [0.19] & [38.2]     & \PsigmaA  \\ 
Gal1        & [3.37]       & [22.33]        & [0.0] & [0.0]          & [0.08] & [27.1]     & \PsigmaC  \\ 
Gal2        & [1.56]       & [25.42]        & [0.081] & [24.1]          & [0.14] & [28.3]     & \PsigmaB  \\ 
Gal3        & [8.37]       & [29.13]        & [0.083] & [71.2]          & [0.09] & [17.1]     & \PsigmaD  \\
\cutinhead{Fixed components}
Halo2       & [7.11]     & [290]        & [0.0] & [0.0]          & [100] & [1500]     & [950]  \\ 
L* galaxy  & \nodata & \nodata & \nodata & \nodata &  [0.15]  &     [40]&  [140]  
\enddata 
 \tablecomments{All coordinates are measured in arcseconds relative to the center of the BCG, at [RA, Dec]=[162.66637 0.285224]. The ellipticity is expressed as $e=(a^2-b^2)/(a^2+b^2)$. $\theta$ is measured north of West. Error bars correspond to 1 $\sigma$ confidence level as inferred from the MCMC optimization. Values in square brackets are for parameters that were not optimized. The location and the ellipticity of the matter clumps associated with the cluster galaxies and the BCG were kept fixed according to their light distribution, and the fixed parameters determined through scaling relations (see text).}
\end{deluxetable*}

All the parameters of the cluster PIEMD halo were allowed to vary except for 
$r_{cut}$, which is not constrainable by the lensing evidence and was thus set to 
1.5 Mpc. We selected cluster member galaxies in the $\sim30'$ SuprimeCam 
field of view, as red-sequence galaxies in a color-magnitude diagram. All 
galaxies within $6.5 \arcmin$ from the BCG and brighter than $i=23$ mag 
were included in the model, with positional parameters ($x$, $y$, $e$, $\theta$) 
that follow their observed measurements, $r_{core}$ fixed at 0.15 pc, and 
$r_{cut}$ and $\sigma$ scaled with their luminosity \citep[for a description of the 
scaling relations see][]{limousin2005}. Four foreground galaxies have a more 
direct influence on the lensed galaxies due to their proximity to the lensed images. 
We thus allowed their velocity dispersions to be solved for in the lens modeling 
process. In particular, we note a galaxy at ($\alpha$,$\delta$) = 10:50:39.704, 
$+$00:17:29.14 -- identified as ``Gal1'' in Figure~\ref{fig:lensmodel} -- that is 
not on the cluster red sequence, but its perturbation of the lensing potential is 
key to the lensing configuration of \arcname. Since the deflection is linear both 
with the distance term and the velocity dispersion, we included this galaxy in 
the same lens plane as the cluster, but note that its best-fit fiducial velocity 
dispersion scales with its dynamical mass and with its distance term. This 
approach is described in more detail by \citet{johnson2014}. We are 
ignoring the second order effects that result from halos residing in different 
lens planes \citep{daloisio2013,mccully2014}, including other projected structure 
which exists along the line-of-sight toward the cluster lens, SDSS J1050+0017 
\citep{bayliss2014a}, but these corrections contribute insignificantly to the 
uncertainty in the magnification derived from our lens modeling of this cluster.

The distribution of cluster galaxies shows a secondary concentration $\sim 290''$ 
north of the BCG. Thus another PIEMD halo is included at the position of the 
brightest galaxy at that position (see Table~\ref{table.lensmodel}). The inclusion 
of that structure is also supported by the weak lensing analysis of \citet{oguri2012}.

The results of preliminary lens models indicated that several parameters are not 
well constrained by the lensing evidence. These include all the parameters 
of the secondary cluster-scale halo (a large range in $\sigma$ is allowed), and 
the scaling relation parameters for cluster member halos; these parameters were 
fixed in subsequent models. Table~\ref{table.lensmodel} lists the best-fit 
parameters and uncertainties, and values of fixed parameters. The model 
uncertainties were determined through the MCMC sampling of the parameter 
space and 1$\sigma$ limits are given. The image plane RMS of the best-fit
lens model is $0\farcs31$. 

The lensing cluster SDSSJ~1050+0017 was first published by \citet{oguri2012}, 
which included a simplified strong lens model. The model was based on constraints 
from only one lensed galaxy (A), and assuming $z=2\pm1$ as a best-guess for its 
redshift, which was not measured at the time. \citet{oguri2012} report an Einstein 
radius for the fiducial arc redshift of 16.1\arcsec~with large error bars due to the 
redshift uncertainty. For the same arc, we find that the Einstein radius 
(defined here as the radius of a circle with the same area as the critical curve) 
is 17.5\arcsec, well in line with the initial model presented by \citet{oguri2012}. 
 
\subsubsection{The Magnification of \arcname}

The lensing magnification depends strongly on the location along the arc. 
Areas in close proximity to the critical curve (plotted as a red line in 
Figure~\ref{fig:lensmodel}) have the highest magnification, and the magnification 
decreases farther from the critical curve. To convert the measurements in this 
paper to their intrinsic, unmagnified values, we calculate the total magnification 
inside the relevant aperture within which each measurement was made. The 
boundaries of the aperture are ray-traced to the source plane; we then measure 
the area covered by the aperture in the image plane, and divide it by the area covered 
by that aperture in the source plane, thus averaging over magnification gradients 
within the aperture. This approach overcomes the problem of pixels very close to 
the critical curve with extremely high magnification artificially driving the average 
magnification to a higher value. 

To derive the uncertainties in the magnification, we compute many models 
with parameters drawn from the MCMC sampling which represent a 1$\sigma$ 
range in the parameter space. The total magnifications are \firemag~
in the aperture used for the FIRE spectroscopy, \gmosmag~ in the GMOS aperture, 
and \photmag~ in the aperture used for the photometric measurement of the arc.

We take the magnification factors that correspond to the slit apertures for the FIRE 
and GMOS spectrum to correct the line flux measurements described in 
$\S$~\ref{sec:lprofiles}. The resulting magnification-corrected (i.e. intrinsic) 
emission line fluxes that result from the brightest knot (observed with FIRE, 
GMOS and MagE) are reported in Table~\ref{tab:fluxes}.

\begin{deluxetable}{lccc}
\tablecaption{Emission Line Flux Measurements\label{tab:fluxes}}
\tablewidth{0pt}
\tabletypesize{\tiny}
\tablehead{
\colhead{} &
\colhead{} &
\colhead{measured flux\tablenotemark{a}  } & 
\colhead{dereddened flux\tablenotemark{b}  } \\
\colhead{ Ion } &
\colhead{ $\lambda_{rest}$} &
\colhead{$\times$10$^{-18}$ } & 
\colhead{$\times$10$^{-18}$ } \\
\colhead{ } &
\colhead{ } &
\colhead{(erg cm$^{-2}$ s$^{-1}$) } & 
\colhead{(erg cm$^{-2}$ s$^{-1}$) } }
\startdata
He {\small II}  & 1640.42 &  0.29$\pm$0.04 & 2.7$^{+1.6}_{-1.1}$   \\
O {\small III}] & 1660.81  & 0.10$\pm$0.04 & 0.9$^{+0.6}_{-0.4}$ \\
O {\small III}] & 1666.15 &  0.29$\pm$0.04 &  2.7$^{+1.6}_{-1.1}$  \\
N {\small III}]\tablenotemark{c} & 1750.71 & 0.08$\pm$0.03 &  0.6$^{+0.3}_{-0.2}$  \\
Si {\small III}] & 1882.47  &  0.26$\pm$0.04 &  2.1$^{+1.1}_{-0.7}$  \\
Si {\small III}] & 1892.03  &  0.15$\pm$0.04 & 1.2$^{+0.7}_{-0.4}$ \\
\ciii\tablenotemark{d}   & 1907.68 &  1.14$\pm$0.07 &  9.0$^{+4.6}_{-3.0}$ \\ 
\ciii\tablenotemark{d}   & 1908.73 &  0.69$\pm$0.08 &  5.5$^{+2.8}_{-1.9}$  \\
O [{\small II}]\tablenotemark{e}   & 3727.09 & 2.26$\pm$0.06 & 8.6$^{+2.6}_{-2.0}$  \\
O [{\small II}]\tablenotemark/{e}   & 3729.88 & 2.22$\pm$0.06 & 8.4$^{+2.6}_{-2.0}$   \\
Ne [{\small III}] &  3870.16 &  1.83$\pm$0.04 & 6.7$^{+1.9}_{-1.6}$  \\
H-$\gamma$ &   4341.68 &  2.96$\pm$0.12 & 9.5$^{+2.5}_{-2.0}$  \\
O [{\small III}] & 4364.44  &  $<$0.81  & $<$3.3  \\
H-$\beta$  &  4862.68 &  5.71$\pm$0.06  &  16$^{+4}_{-3}$   \\
O [{\small III}]  & 4960.29  &  13.44$\pm$0.10 &  38$^{+8}_{-7}$  \\
O [{\small III}] &  5008.24  &  46.12$\pm$0.12  & 127$^{+29}_{-23}$ 
\enddata
\tablenotetext{a}{~Values reported here are corrected for the lensing magnification.}
\tablenotetext{b}{~These are the magnification-corrected line fluxes after applying a 
reddening correction using a \citet{calzetti2000} extinction law with A$_{V} =$ 1, with 
reported uncertainties that include the uncertainty in the reddening correction.}
\tablenotetext{c}{~This line is only marginally detected at $\sim$2$\sigma$.}
\tablenotetext{d}{~These lines are blended in the GMOS spectra; we extract the 
individual line fluxes by fitting a double gaussian model to the combined line profile, 
see $\S$~\ref{sec:electrondensity}.}
\tablenotetext{e}{~These lines are blended in the FIRE spectra; we extract the 
individual line fluxes by fitting a double gaussian model to the combined line profile, 
see $\S$~\ref{sec:electrondensity}.}
\end{deluxetable}

\section{Recovering Physical Quantities}
\label{sec:analysis}

In this section we constrain various different physical properties of the lensed source, 
\arcname. The following subsections describe our methods for recovering parameter 
values, and the resulting values are shown in Table~\ref{tab:logoh} (stellar mass, extinction, 
star formation rate, electron density \& temperature, and ionization parameter) and 
Table~\ref{tab:abundances} (metallicity and relative ionic abundances).

\subsection{Broadband SED Fitting}

We model the observed spectral energy distribution using the fitting 
code FAST \citep{kriek2009} at fixed spectroscopic redshift with 
\citet{bruzchar2003} stellar population synthesis models, a 
\citet{chabrier2003} IMF and \citet{calzetti2000} dust extinction law. 
We adopt exponentially decreasing star formation histories with 
minimum e-folding time log($\tau$) $=$ 8.5 yrs \citep[e.g., ][]{swuyts2011} 
and the metallicity is allowed to vary from 0.2Z$_{\odot}$ to Z$_{\odot}$. 
This results in a stellar mass estimate log(M$_{*}$/M$_{\odot}$) $=$ 11.0 
$\pm$ 0.15 (statistical) $\pm$ 0.2 (systematic) M$_{\odot}$. We combine 
this with the magnification acting on the arc as computed from the strong 
lens models described above ($\mu =$ \photmag) to recover 
the intrinsic stellar mass of \arcname: log(M$_{*}$/M$_{\odot}$) $=$ 
9.5 $\pm$ 0.15 (statistical) $\pm$ 0.2 (systematic) 

Because of the relatively large point spread function (PSF) of the IRAC bands 
there is possible contamination in the measured flux of \arcname~in the 
IRAC 3.6 and 4.5$\mu$m bands from a nearby foreground galaxy 
that is separated from the arc by $\sim$1.5\arcsec. As a test we have explored 
SED fits with the IRAC flux reduced by a factor of 2$\times$ (an extreme case); 
we find that this factor of 2 contamination case only weakly affects the best-fit 
stellar mass, and that the possible contamination is sub-dominant to other 
systematic uncertainties in the SED-derived stellar mass 
\citep{shapley2005,swuyts2011,conroy2009,conroy2010a,conroy2010b}.

\subsection{Reddening Constraints from Balmer Lines}

The H-$\beta$ and H-$\gamma$ Balmer lines are detected in the FIRE spectra, 
which allows us to place a constraint on the reddening due to interstellar dust in 
the rest-frame. H-$\beta$ is well-detected in the FIRE data, but unfortunately the 
H-$\gamma$ line falls in a region of poor atmospheric throughput and on top of 
sky lines, which degrades our ability to precisely measure the line flux. 
\arcname~appears from all indications to be a relatively low-mass galaxy with 
active ongoing star formation, so we assume a \citet{calzetti2000} extinction 
law. The measured H-$\beta$/$\gamma$ ratio indicates an internal extinction of 
E(B-V) $=$ 0.14$^{+0.38}_{-0.14}$, which is effectively an upper limit of  E(B-V) 
$<$ 0.52. This agrees with the results of the best-fit SED model, which prefers 
A$_{V} =$ 1.0 $\pm$ 0.2. From here on out we proceed with an extinction value 
of A$_{V} =$ 1.0 and the \citet{calzetti2000} extinction law and correct all reddening 
sensitive measurements accordingly. Achieving better 
constraints on the internal reddening will be challenging due to H-$\alpha$ being 
redshifted out of the K band, and H-$\gamma$ falling into a region of poor 
atmospheric transmission.

\subsection{Damped Lyman-$\alpha$ Absorption}

From the GMOS and MagE spectra it also appears as though \arcname~is a damped 
Lyman-$\alpha$ Absorption (DLA) system. The GMOS spectrum is higher S/N and we 
use it to fit a Voigt profile using the XIDL procedure \texttt{x\_fitdla}. Even in the GMOS data 
the S/N is falling off rapidly in the region of the damped Ly-$\alpha$ absorption due to the 
combination of 1) the throughput of the grating decreasing at bluer wavelengths and 2) 
the continuum emission being suppressed by the Ly-$\alpha$ absorption. We use 
$\Delta \lambda \sim $5600-6200\AA -- corresponding to $\Delta \lambda 
\sim$1210-1340\AA~in the rest frame -- to fit the DLA profile (see 
Figure~\ref{fig:gmosspec}). In this region of the spectrum the data fall off from S/N 
per spectral pixel of $\sim$6 at 6200\AA~to $<$1 at 5600\AA. The low S/N limits our 
ability to precisely constrain the centroid of the DLA profile, but we 
find that the data prefer a high column density, log(N$_{HI}) >$ 21.5 cm$^{-2}$, independent 
of the precise redshift centroid of the DLA feature. The width of the velocity broadening 
component of the profile is also unconstrained. Higher S/N observations of the spectrum 
blueward of 6000\AA~will be necessary to make a precise measurement of the DLA 
feature, but the available data strongly indicate the presence of a large quantity of 
neutral hydrogen.

\subsection{Star Formation Rate Estimates}


Following the calibrations of \citet{kennicutt1998}, we compute the star formation rate 
(SFR) from the FIRE observations of the brightest knot of \arcname~ using the 
H-$\beta$ and [O {\small II}]$\lambda$$\lambda$3727,3729 emission lines. Both the 
H-$\beta$ and [O {\small II}]$\lambda$$\lambda$3727,3729 SFR estimates measure 
the instantaneous star formation, because 
they probe the integrated luminosity of massive stars blueward of the Lyman limit 
(i.e. ionizing photons). We correct these SFR measurements by using the strong 
lens model described in $\S$~\ref{sec:slmodel} to compute the magnification factor 
that applies to the region of the arc covered by the FIRE slit. In this case, the FIRE slit 
measures a region of the arc with a total magnification, $\mu =$ \firemag, 
and after correcting for this magnification we compute the SFR (for the knot covered by 
FIRE only) of SFR$_{H\beta} =$ 84 $\pm$ 24 M$_{\odot}$ yr$^{-1}$ and 
SFR$_{O II} =$ 55 $\pm$ 25 M$_{\odot}$ yr$^{-1}$, where the large uncertainties 
reflect a 20\% uncertainties in the absolute flux calibration of the FIRE data, as well 
as measurement errors and a 30\% scatter in the calibration between and 
luminosity in the case of SFR$_{O II}$. 

\subsection{Electron Density}
\label{sec:electrondensity}

Our spectra include three distinct doublet lines that provide a measurement of the 
electron density, n$_{e}$, in the HII regions that are responsible for the observed 
nebular line emission \citep{osterbrock1989}. The GMOS spectrum contains both 
\ciii$\lambda$$\lambda$1907,1909 and \siiii$\lambda$$\lambda$1882,1893, where 
the Si lines are well-separated and the \ciii~lines are blended but resolved. The FIRE 
spectrum resolves the [O {\small II}] $\lambda$$\lambda$3727,3729 lines. All of these 
line pairs are located very close to one another in wavelength so that the uncertainty 
regarding the internal reddening does not factor into the electron density determination. 
We use the curves from \citet{osterbrock1989} to convert line ratios into electron 
density; the reported uncertainties are those associated with the measurement 
uncertainties in the line ratios and assume no additional systematic uncertainties in 
the line ratio vs. n$_{e}$ curves of \citet{osterbrock1989}. We compute initial estimates 
of the electron density beginning with an assumed electron temperature, T$_{e} =$ 
10,000 K, and then iterate the computations described in this Section and the 
following section to arrive at the converged values presented here.

\begin{figure}
\centering
\includegraphics[scale=0.45]{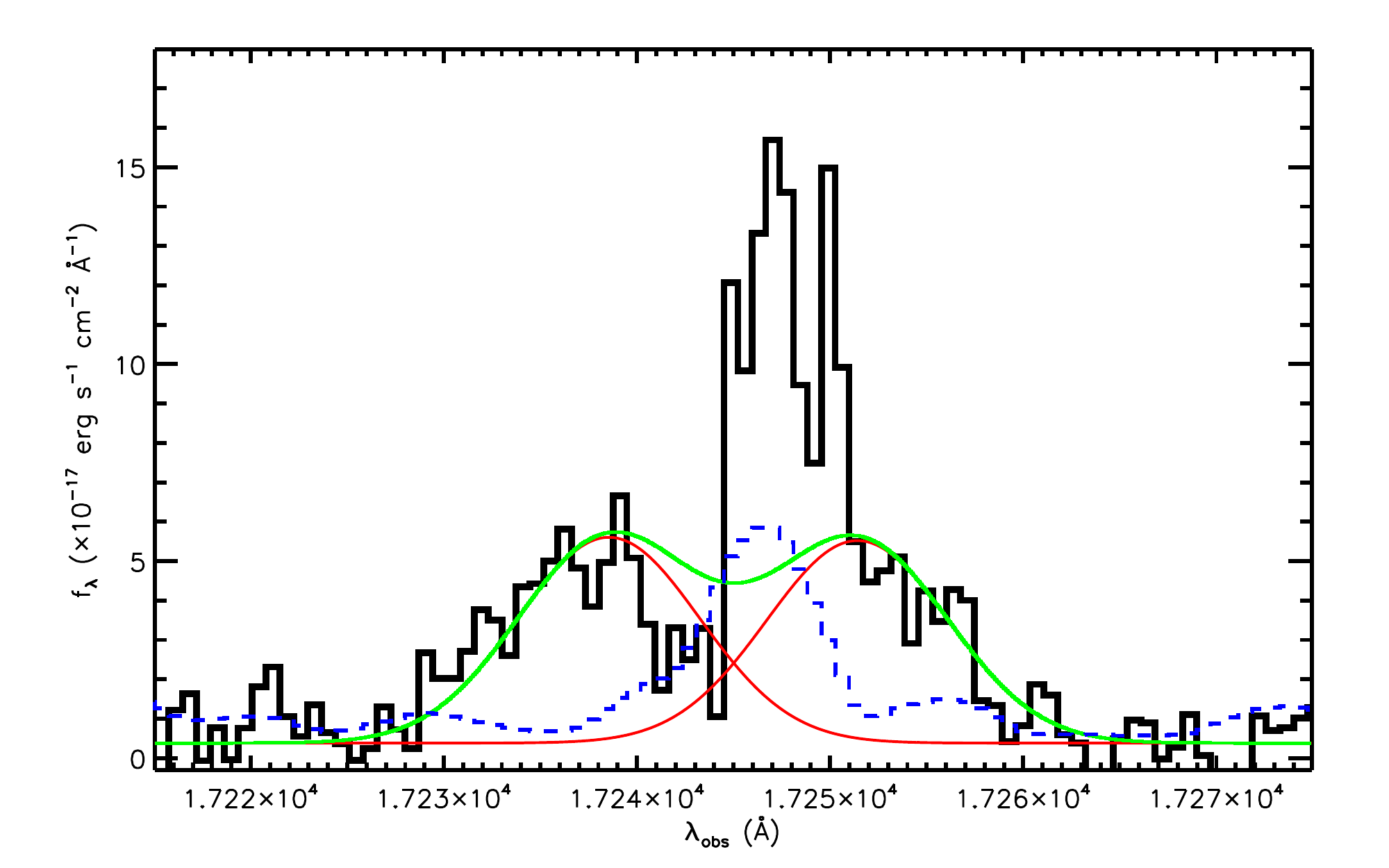}
\includegraphics[scale=0.45]{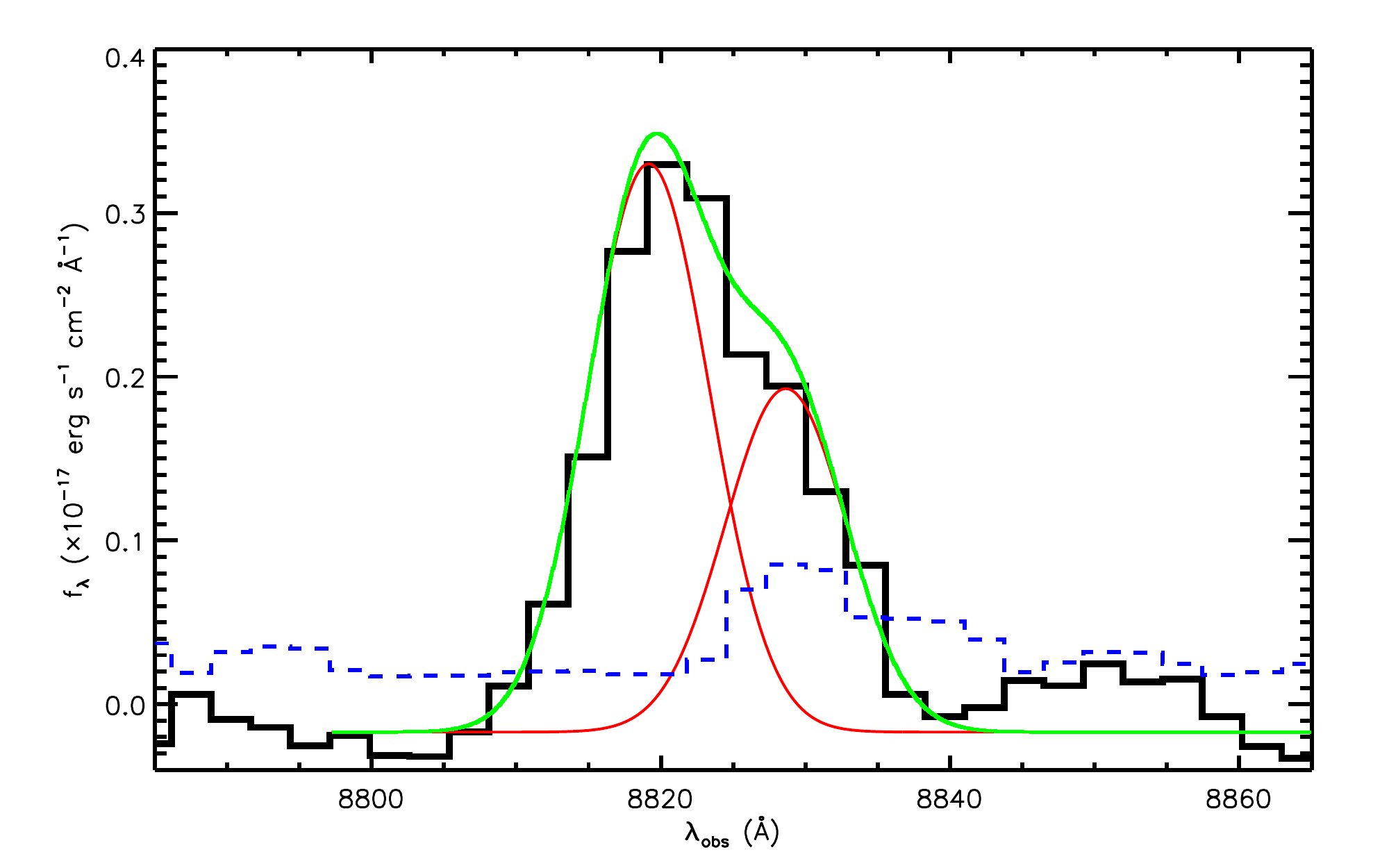}
\includegraphics[scale=0.45]{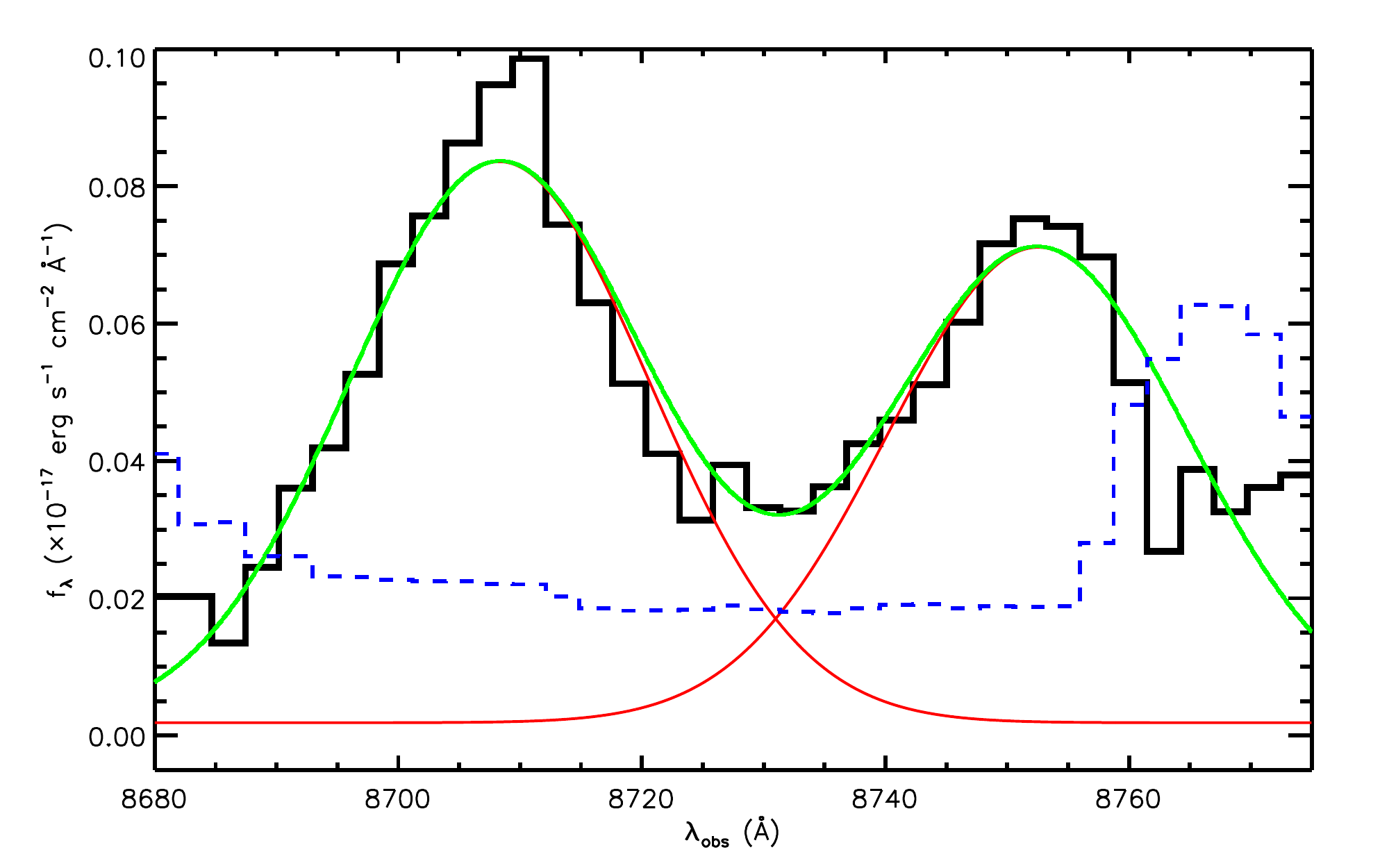}
\caption{\scriptsize{
{\it Top}: The double gaussian fit to the [O II] $\lambda$$\lambda$3727,3729 doublet. 
The FIRE spectrum 
is plotted as the black solid histogram, with the uncertainty plotted as the blue dashed 
histogram. Over-plotted in are the individual gaussian fits (red curves) to the two [O II] 
lines, along with the sum of the two gaussians (green curve). These lines are coincident with a 
bright sky line, which appears to have produced an artificial deficit of flux blueward of the 
centroid of the sky line and an excess redward. The double gaussian fit allows us to 
recover a good estimate of the true line shapes and strengths of the [O II] doublet.
{\it Middle}: The double gaussian fit to the \ciii$\lambda$$\lambda$1907/1909 doublet. 
Again, the GMOS spectrum is plotted in black with the uncertainty in blue (and dashed). 
Individual (red curves) gaussians for the separate lines, as well as the summed double 
gaussian profile (green curve) are over-plotted.
{\it Bottom}: The double gaussian fit to the Si {\small III}]$\lambda$$\lambda$1882,1892 
doublet using the same line/color scheme as the middle panel. The Si {small III} lines 
are well-separated in our spectra (though the lines have much lower S/N than the [O II] 
and \ciii~doublets).
}}
\label{2gaussfits}
\end{figure}


The cleanest lines that we can measure in the available data are 
\siiii$\lambda$$\lambda$1882,1892, 
which are well-separated in wavelength and unaffected by strong sky subtraction 
residuals. We also measure the line ratios for \ciii~and [O {\small II}], though systematic 
uncertainties from sky lines make them both less robust measurements than the 
\siiii~ratio. The \ciii~and \siiii~lines both probe a relatively higher density range and 
therefore cannot precisely constrain n$_{e}$ values below $\sim$ 10$^{3}$ cm$^{-2}$; 
the [O {\small II}]$\lambda$$\lambda$3727,2729 line ratio provides constraints below 
n$_{e} \sim$ 10$^{3}$ cm$^{-2}$ but are unfortunately limited by a bright sky line 
residual in our data. The specific n$_{e}$ constraints from are data are summarized 
as follows:

\begin{enumerate}

	\item {\bf Si {\small III}]$\lambda$$\lambda$1882,1892 ---} We measure this line ratio 
	to be 1.6 $\pm$ 0.2 in the GMOS spectrum. This value corresponds to an electron 
	density n$_{e} =$ 10$^{3}$ cm$^{-2}$ for nebular line emitting regions 
	within \arcname. Incorporating the uncertainty in the Si {\small III}] line ratio results 
	in an upper limit,  n$_{e} \leq$ 2 $\times$ 10$^{3}$ cm$^{-2}$. This is fully consistent 
	with \arcname~being in the low-density regime, though the \siiii~doublet ratio 
	transitions at higher densities and therefore 
	does not provide as powerful a constraint on low density values of n$_{e}$ as as 
	well-measured [O {\small II}] 3727,2729 line ratio would.

	\item {\bf [O {\small II}]$\lambda$$\lambda$3727,2729 ---} The measurement of 
	[O {\small II}]$\lambda$$\lambda$3727,3729 
	is complicated by the unfortunate fact that the lines are redshifted onto the location 
	of a bright sky line in the NIR. In order to attempt to recover the line ratio from the 
	FIRE data we fit a double gaussian profile to the resolved 
	[O {\small II}]$\lambda$$\lambda$3727,3729 
	lines, holding the redshift fixed to the values taken from the mean of the other 
	nebular emission lines apparent in the FIRE data (see Table~\ref{zsystemic}). This 
	line profile fitting method allows us to recover an estimate of the true line strengths, 
	ignoring contamination from the bright sky line residuals. We measure a line ratio 
	of 1.0 $\pm$ 0.4; the central value implies an electron density of n$_{e} \sim$ 2-3$\times
	$10$^{2}$ cm$^{-2}$, though the large uncertainty effectively encompasses all physically 
	plausible density values.

	\item {\bf C {\small III}]$\lambda$$\lambda$1907,1909 ---} The 
	\ciii$\lambda$$\lambda$1907,1909 line ratio is fit in the same way as 
	the [O {\small II}]$\lambda$$\lambda$3727,3729, and we find a line ratio of 1.65 
	$\pm$ 0.14, which is non-physical but less than 1$\sigma$ from the maximum 
	value for the line ratio in the low-density limit ($\sim$1.55), and implies a 
	2$\sigma$ limit of n$_{e} \lesssim$ 3$\times$10$^{3}$ cm$^{-2}$. We also note 
	that the line measurements for \ciii~suffer from a similar source of uncertainty as the 
	O {\small II}]$\lambda$$\lambda$3727,2729 doublet due to the lines being redshifted 
	to fall on to a bright sky line. However the nod-and-shuffle strategy employed with 
	the GMOS data helps significantly in limiting the sky line subtraction uncertainties to 
	the poisson minimum.

\end{enumerate}

The two gaussian profile fits to each of the [O {\small II}], C {\small III}], and Si {\small III}] 
doublets are shown in Figure~\ref{2gaussfits}. From the three electron density indicators, 
there is a strong preference for the low-density regime, with n$_{e} \leq $10$^{3}$ cm$^{-2}$. 
We also note that the apparent offset in density values preferred by the [O {\small II}] and 
C {\small III}] doublets may simply reflect the fact that these doublets trace the electron 
density of the low and medium ionization zones, respectively 
\citep{quider2009,christensen2012b,james2014}.

\subsection{Electron Temperature}

There is very strong [O {\small III}]$\lambda$$\lambda$4960,5008 emission, but no 
detection of [O {\small III}]$\lambda$4364 in the FIRE spectrum. We use the non-detection 
of [O {\small III}]$\lambda$4364 to place a limit on the electron temperature in the 
nebular line emitting regions following the method of \citet{izotov2006} (Section 3.1, 
Equations 2 \& 3). \citet{izotov2006} point out that the electron density becomes 
unimportant to the electron temperature determination for values of n$_{e} \lesssim 
10^{3}$ cm$^{-2}$, and our constraint on n$_{e}$ from the previous section places 
\arcname~in that regime. From the [O {\small III}]$\lambda$$\lambda$4960,5008 
lines and the [O {\small III}]$\lambda$4364 limit we place a limit on T$_{e}$ (O {\small III}) 
$\leq$ 14,000 K.

An alternative method for estimating the electron temperature is available to us by 
measuring the flux ratios between the \oiii$\lambda$$\lambda$1661,1666 and 
[O {\small III}]$\lambda$$\lambda$4960,5008 \citep{villar-mart2004,erb2010}. We 
have good detections of all the relevant lines, so this method allows us to measure a 
precise temperature rather than a limit. However, this measurement has its own 
caveats, primarily a large sensitivity to the intrinsic extinction, and a large uncertainty 
between the absolute flux calibrations of the optical (GMOS) and NIR (FIRE) spectra. 
With these caveats in mind, we measure T$_{e}$ = 11300$^{+1400}_{-1000}$ K 
(measurement uncertainties only), which agrees well with the T$_{e}$ constraint from 
the FIRE spectrum alone. When we fold in the additional uncertainty between the relative 
flux calibrations of the GMOS and FIRE data, as well as the uncertainty in our best-fit 
value for the dust extinction (A$_{V} = 1 \pm 0.2$) we find that T$_{e}$ constraint from 
the \oiii$\lambda$$\lambda$1661,1666 to [O {\small III}]$\lambda$$\lambda$4960,5008 
line ratio is quite broad: 
4000 $<$ T$_{e}$ $<$ 15000 K. This is less 
constraining (for physically plausible values of T$_{e}$) than the limit derived from the 
non-detection of [O {\small III}]$\lambda$4364. From here on out we proceed with the 
T$_{e}$ $\leq$ 1.4 $\times$ 10$^{4}$ K constraint on the electron temperature.

\begin{deluxetable}{ll}[t]
\tablecaption{Physical Parameter Constraints\label{tab:logoh}}
\tablewidth{0pt}
\tabletypesize{\tiny}
\tablehead{
\colhead{Physical} &
\colhead{Value} \\
\colhead{Quantity } &
\colhead{ } } 
\startdata
log(M$_{*}$/M$_{\odot}$)  & 9.5 $\pm$ 0.35  \\
A$_{V}$   &  1.0 $\pm$ 0.2  \\
N$_{HI}$ (cm$^{-2}$)   &  $\geq$ 10$^{21.5}$ \\
SFR$_{H\beta}$ (M$_{\odot}$ yr$^{-1}$) & 84 $\pm$ 24 \\
SFR$_{OII}$ (M$_{\odot}$ yr$^{-1}$) & 55 $\pm$ 25   \\
n$_{e}$ (cm$^{-3}$) &  $\leq$ 10$^{3}$ \\
T$_{e}$ ($\lambda$4364) (K) &  $<$ 14000  \\
T$_{e}$ ($\lambda$$\lambda$1666) (K) &  $<$ 15000 \\
log(U) (O3O2)   &   $-$2.22 $\pm$ 0.15  \\
log(U) (Ne3O2)   &   $-$2.7$^{+0.3}_{-0.2}$  
\enddata
\end{deluxetable}

\subsection{Oxygen Abundance Indicators}
\label{sec:oabundance}

In the the next two subsections we apply several different Oxygen and ionic 
abundance indicators  to our observations of \arcname. The results are summarized 
in Table~\ref{tab:abundances}. For all of these calculations we include a systematic 
uncertainty term that results from the uncertainty in the extinction correction 
(A$_{V} = 1 \pm 0.2$) that is applied to the Oxygen lines.

\subsubsection{T$_{e}$ Direct Oxygen Metallicity}
From the detected oxygen lines we can use the direct T$_{e}$ 
method to constrain the metallicity of \arcname~using the prescription outlined 
by \citet{izotov2006}. This metallicity measurement accounts 
for the singly (O$^{+}$/H$^{+}$) and doubly ionized (O$^{++}$/H$^{+}$) oxygen 
atoms, but ignores triply ionized oxygen which only contributes significantly to the 
measured abundance in regions of extremely high ionization. The metallicity 
constraint from this method is \logoh$_{Te} \geq$ 8.05 ($\geq$ 0.22Z$_{\odot}$) when 
we use the upper limit on the electron temperature from the 
[O {\small III}]$\lambda$$\lambda$4960,5008 to [O {\small III}]$\lambda$4364 ratio. 
	
\subsubsection{R$_{23}$}
The R$_{23}$ index measures the strength of [O {\small II}] and 
[O {\small III}] lines against Balmer line emission from hydrogen: 
log([$\lambda$3727$+\lambda$3729$+\lambda$4960$+\lambda$5008]/H-$\beta$). 
This index is problematic because it is double valued, with both high metallicity and low 
metallicity branches. We measure log(R$_{23}$) $=$ 1.05 $\pm$ 0.06. The high value 
for R$_{23}$ places \arcname~in the transition zone between the upper and lower 
branches of the observed R$_{23}$ -- metallicity relation. There are several different 
calibrations for the R$_{23}$ index in the literature for one or both branches; we 
compute the R$_{23}$ based \logoh~for these different calibrations.

The upper branch calibration of \citet{zaritsky1994} yields \logoh$_{ZKH94} =$ 8.30 $\pm$ 
0.09, or 0.4Z$_{\odot}$, in good agreement with both the limit and the measurement 
from the T$_{e}$ method. Using the calibration of \citet{pilyugin2005} we get an 
upper branch abundance of \logoh$_{P05,upp} =$ 8.19 $\pm$ 0.11 and a lower branch 
value of \logoh$_{P05,low} =$ 8.16 $\pm$ 0.11. Finally, we also apply the R$_{23}$ 
calibration of \citet{kobulnicky2004}, which yields an upper branch value of 
\logoh$_{KK04,upp} =$ 8.49 $\pm$ 0.11 and a lower branch value of 
\logoh$_{KK04,low} =$ 8.26 $\pm$ 0.11.

\subsubsection{Ne3O2}
The ratio of [Ne~{\small III}]$\lambda$3869 to 
[O {\small II}]$\lambda$$\lambda$3727,3729 has also been calibrated as an indicator 
of the log(O/H) metallicity by \citet{shi2007}. We measure log(3869/3727$+$3729) 
$=$ $-$0.41, which corresponds to \logoh~$=$ 7.5 $\pm$ 0.2. The error bar 
reported here does not include the uncertainty in the zero point of the Ne3O2 
calibration, which is quite large ($\sim$ 0.7 dex), and while we report the Ne3O2 
metallicity estimate in the spirit of being thorough, we do not use it to compute the 
mean metallicity of \arcname~(see Table~\ref{tab:abundances}).

\subsubsection{Average Oxygen Abundance Metallicity}

In addition to the results of each of the individual Oxygen abundance 
metallicity indicators, we also show in Table~\ref{tab:abundances} the 
average metallicity value of all of the Oxygen abundance indicators calculated in the 
previous sections, excluding the Ne3O2 method due to the extremely large uncertainty 
in the zero point of the Ne3O2 metallicity diagnostic.

\subsection{Ionic Abundance Ratios}
\label{sec:ionabundance}

\subsubsection{Ionization Parameter, log(U)}

From the available data we can follow \citet{kewley2002} and \citet{kobulnicky2004} to 
estimate the ionization parameter from the ratio of [O {\small III}] to [O {\small II}] line 
fluxes. This is computed iteratively with the \citet{kobulnicky2004} oxygen abundance 
computed in $\S$~\ref{sec:oabundance}, above. The preferred ionization parameter 
from the rest-frame optical emission lines is log(U) $=$ $-$2.22 $\pm$ 0.15. This ionization 
parameter is not nearly so extreme as has been found in a few extreme star bursting 
galaxies at z $\sim$2-3.3, where log(U) is measured to be as high as $\sim -1$ 
\citep[e.g.;][]{villar-mart2004,erb2010}. We also use the 
Ne3O2 line ratio diagnostic for recovering the ionization parameter as recently 
proposed by \citet{levesque2014}, which we agree is likely a much better use of 
this line ratio measurement than the Ne3O2 metallicity indicator. Ne3O2 yields an 
ionization parameter estimate of log(U) $=$ $-$2.7$^{+0.3}_{-0.2}$. The uncertainty 
here includes contributions from the line flux measurements, the reddening correction, 
and the $\sim 0.1$dex uncertainty in the average metallicity measurement (see 
Table~\ref{tab:abundances}).

It is not clear which of the two diagnostics above provides the more reliable estimate 
of log(U). There are reasons to suspect, however, that both of the preceding ionization 
parameter estimates are unreliable, or at least incomplete in their description of the 
physical conditions within \arcname. Specifically, the detection of strong 
He~{\small II}$\lambda$1640 emission, and significant excess emission in the P 
Cygni line profiles of Si {\small IV} and C {\small IV} both prefer larger values of log(U). 
We discuss these features in more detail in $\S$~\ref{sec:heii} and $\S$~\ref{sec:pcygni}.

\begin{deluxetable}{ll}[t]
\tablecaption{Ioni Abundance Measurements\label{tab:abundances}}
\tablewidth{0pt}
\tabletypesize{\tiny}
\tablehead{
\colhead{Ion Abundance Ratio} &
\colhead{Value} 
}
\startdata
\logoh~(T$_{e}$ Direct)........................ & $>$8.05  \\
\logoh~(R$_{23,ZKH94}$).....................     & 8.30 $\pm$ 0.09  \\
\logoh~(R$_{23,P05-upp}$)...................     & 8.19 $\pm$ 0.11  \\
\logoh~(R$_{23,P05-low}$)...................     &  8.16 $\pm$ 0.11  \\
\logoh~(R$_{23,KK05-upp}$)................     & 8.49 $\pm$ 0.10   \\
\logoh~(R$_{23,KK05-low}$)................     &  8.26 $\pm$ 0.10  \\
\logoh~(Ne3O2)............................ & 7.5 $\pm$ 0.2 ($\pm$ 0.7)\tablenotemark{a}  \\ 
 $<$\logoh$>$ (excl. Ne3O3).............  & 8.28 $\pm$ 0.13  \\ 
 $<$Z/Z$_{\odot}$$>$....................................................  & 0.39$^{+0.13}_{-0.10}$  \\  \\
12$+$log(Ne$^{++}$/H$^{+}$) (Ne$^{++}$/H$^{+}$)................ & 7.13 $\pm$ 0.23 \\ \\
log(C$^{++}$/O$^{++}$) ($\lambda$1909/$\lambda$1666)................. &  $-$0.79 $\pm$ 0.06  \\
log(C$^{++}$/O$^{++}$) ($\lambda$1909/$\lambda$5008)................. &  $-$0.77$\pm$ 0.32  \\ \\
log(N$^{++}$/O$^{++}$) ($\lambda$1750/$\lambda$1666)................. &  $-$1.2 $\pm$ 0.3  \\ \\
log(Si$^{++}$/C$^{++}$) ($\lambda$1892/$\lambda$1909)................. &  $-$1.2 $\pm$ 0.3 \\
log(Si$^{++}$/O$^{++}$) ($\lambda$1892/$\lambda$1666)................. &  $-$2.0 $\pm$ 0.4 
\enddata
\tablenotetext{a}{~The parenthetical uncertainty is the scatter in the zero point for the Ne3O2 
log(O/H) indicator.}
\end{deluxetable}

\subsubsection{Ne$^{++}$ Abundance}

From the strong [Ne {\small III}]$\lambda$3869 emission line we can compute the 
abundance of Ne$^{++}$/H$^{+}$ using equation 7 from \citet{izotov2006}. We find 
12 $+$ log({\small Ne$^{++}$/H$^{+}$}) $=$ 7.13 $\pm$ 0.23. In the Lynx arc, 
\citet{villar-mart2004} approximate that all of the nebular Ne atoms are doubly 
ionized, so that Ne/H $\sim$ Ne$^{++}$/H$^{+}$, but it's not clear that we can 
make the same assumption given that we measure an [O {\small II}] to [O {\small III}] 
ratio that is consistent with an ionization parameter that is considerably lower than the 
log(U) $\sim$ $-$1 that is found in the Lynx arc. We can at least note the lack of 
observable flux from known Ne {\small IV} and Ne {\small V} lines in both the 
rest-UV and rest-optical, which qualitatively argues against these states 
representing a large fraction of the total Ne.

\begin{figure*}[t]
\centering
\includegraphics[scale=0.657]{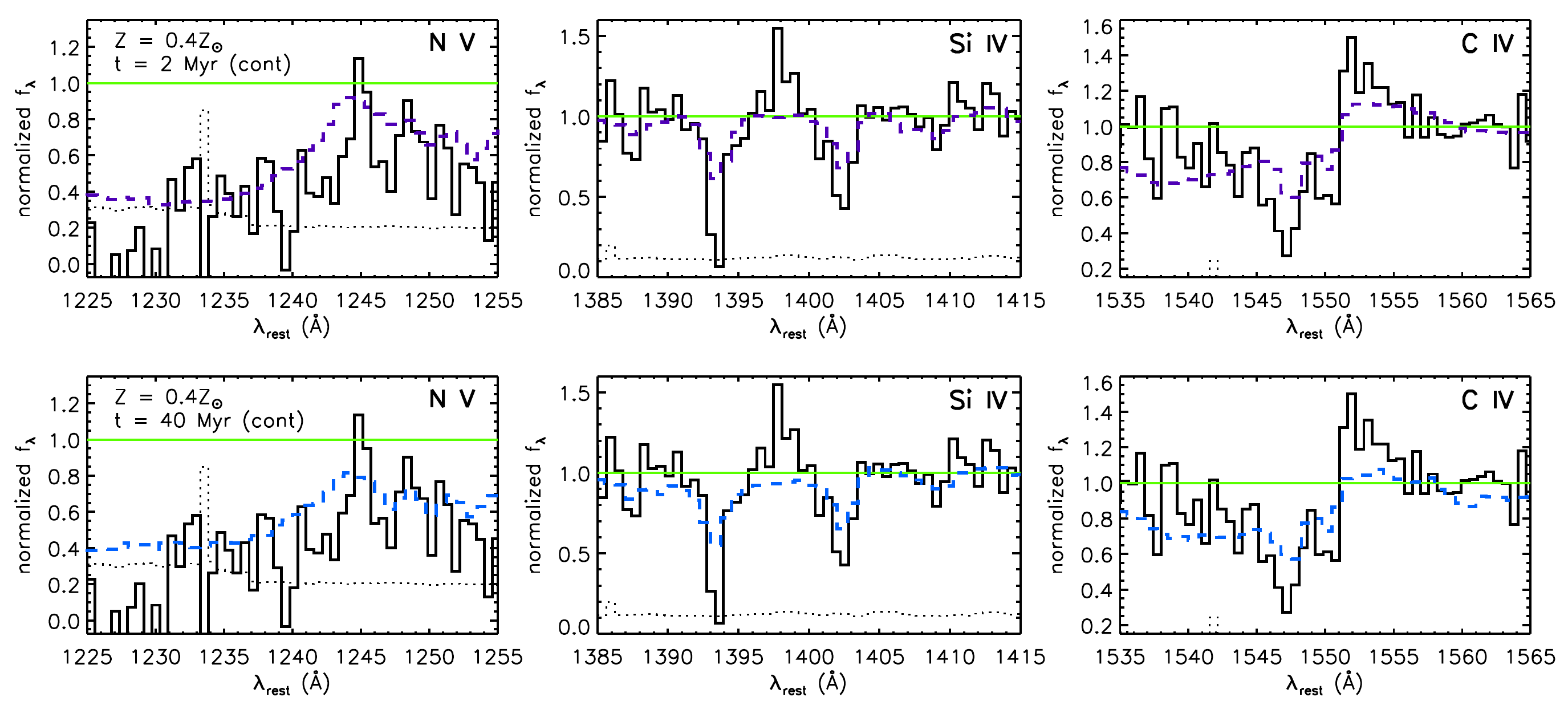}
\caption{\scriptsize{
Comparison of continuous star formation \texttt{S99} models against the P Cygni 
features in the GMOS spectrum. The GMOS data are plotted as solid histograms, 
with the error array over plotted as a dotted line and the fiducial continuum level 
indicated by the horizontal green line. The \texttt{S99} models based 
on empirical LMC/SMC stellar atmospheres are over plotted as the color dashed lines. 
Each row is a different \texttt{S99} model: the ages of the stellar populations are varied 
between the top and bottom rows.}}
\label{fig:pcygni}
\end{figure*}

\subsubsection{C$^{++}$/O$^{++}$ Abundance Ratio}

\citet{garnett1995b} used {\it HST} spectroscopy of dwarf galaxies to measure the 
relative abundances of C$^{++}$/O$^{++}$ and N$^{++}$/O$^{++}$ from rest-UV 
emission lines. We detect the same families of lines in the GMOS spectrum of 
\arcname~and can therefore measure the ion abundance ratios of this galaxy at 
z $=$ 3.6252. We use the ratio of the line intensities of 
\oiii$\lambda$$\lambda$1661,1666 and \ciii$\lambda$$\lambda$1907,1909 
\citep{garnett1995b} to measure log(C$^{++}$/O$^{++}$) $=$ $-$0.79 $\pm$ 0.06. 
This abundance measurement is relatively insensitive to extinction due to the 
similar wavelengths of the lines.

\citet{kobulnicky1998} have also 
shown that it is possible to measure the C$^{++}$/O$^{++}$ from the relative line 
strengths of \ciii$\lambda$$\lambda$1907,1909 and 
[O {\small III}]$\lambda$$\lambda$4960,5008. From this method we measure 
log(C$^{++}$/O$^{++}$) $=$ $-$0.77 $\pm$ 0.32, where the larger 
uncertainty is driven by the uncertainty in the reddening correction due to the fact 
that the lines used with this method span a large range in wavelength.  
Comparing the two C$^{++}$/O$^{++}$ measurements 
we see remarkable agreement, which could be interpreted as a confirmation that 
A$_{V} = 1$ is, in fact, an appropriate extinction value for \arcname. However, the 
values reported here do not reflect the uncertainty that results from the lack of a 
precise electron temperature constraint, and it is possible that the agreement 
results in part from different sources of error fortuitously canceling out.

Previous investigations have shown that there is no appreciable ionization correction 
factor (ICF) necessary for comparing C/O ion ratio given log(U) values in line with what 
we find above for \arcname~\citep{garnett1995b,erb2010}. If our estimate of log(U) is 
correct then the log(C$^{++}$/O$^{++}$) measurements above are effectively telling 
us the total log(C/O) elemental abundance ratios in \arcname.


\subsubsection{N$^{++}$/O$^{++}$ Abundance Ratio}

Similar to the C/O ratio measurement, \citet{garnett1995b} use the ratio of the 
\oiii$\lambda$$\lambda$1661,1666 lines and the N {\small III}]$\lambda$1750 
multiplet to measure the N/O abundance. Nitrogen 
has ionization potentials that are closer to oxygen than carbon, and so an ionization 
correction factor should also not be necessary for inferring the total N/O ratio using 
the N$^{++}$/C$^{++}$ ion ratio.
Applying this method we measure log(N$^{++}$/O$^{++}$) $=$ $-$1.6 $\pm$ 0.2, which 
assuming a negligible ICF, provides an estimate of the relative C/N enrichment: 
log(C/N) $=$ 0.8 $\pm$ 0.2.

\subsubsection{Si$^{++}$/C$^{++}$ and Si$^{++}$/O$^{++}$ Abundance Ratios}

\citet{garnett1995a} demonstrate how the Si/O relative abundance can be computed 
from the rest-UV Si {\small III}]$\lambda$$\lambda$1882,1892 and 
\ciii$\lambda$$\lambda$1907,1909 lines. As shown by \citet{garnett1995a}, the ICF for using 
the ratio of Si$^{++}$ to C$^{++}$ as an approximation of the total Si/C abundance 
is somewhat sensitive to the ionization parameter. Our estimate of log(U) for 
\arcname~implies a fraction of doubly ionized oxygen of X(O$^{++}$) $\sim$ 0.8-0.85 
based on the models from \citet{erb2010}. From \citet{garnett1995a} this implies 
X(Si$^{++}$)X(C$^{++}$) $\sim$0.65-0.9.
This abundance estimate has a 25\% uncertainty from the ICF alone; we estimate it 
to be in the range $\sim$1.1-1.5 with a central value of 1.4 
\citep[see, e.g.;][]{kobulnicky1998}. We find log(Si$^{++}$/C$^{++}$) $=$ $-$1.2 $\pm$ 0.3. 
Assuming our assumptions about the ICF are reasonable, this measurement can be 
combined with our measurements of the log(C/O) abundances to yield 
log(Si$^{++}$/O$^{++}$) $=$ $-$2.0 $\pm$ 0.4.

\subsection{Starburst99 Model Comparisons}
\label{sec:s99}

\begin{figure*}[t]
\centering
\includegraphics[scale=0.657]{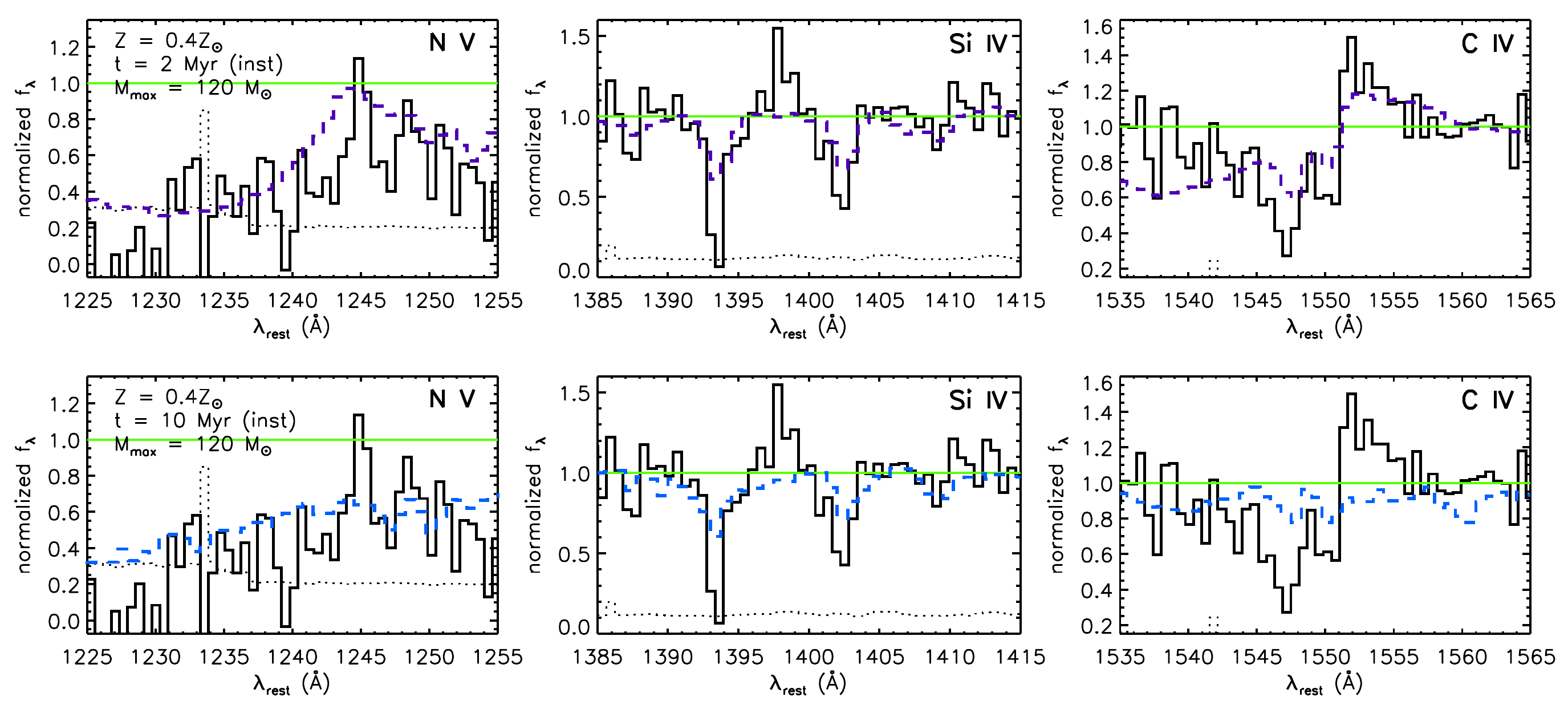}
\caption{\scriptsize{
\texttt{S99} models created assuming a single instantaneous burst of star formation 
rather than the continuous star formation models. The panels are similar to those 
plotted in Figure~\ref{fig:pcygni}, with the top and bottom rows showing stellar populations 
of two different ages. These models allow for stars up to 120M$_{\odot}$ to form, which 
results in strong C IV emission at early times (see the top row of panels) but still does 
not reproduce the combination of absorption and emission P Cygni features that we 
observe in \arcname.}}
\label{fig:pcygni2}
\end{figure*}

The P Cygni ISM absorption/emission features that we observe in the rest-UV are 
the byproduct of powerful winds that are typically associated with Wolf-Rayet (W-R) 
stars. To try and understand the physical implications of the wind features in the 
spectrum of \arcname, we compare our data against \texttt{Starburst99} models 
\citep[\texttt{S99};][]{leitherer2010}.

Both our GMOS and MagE data include each of the 
N {\small V}$\lambda$$\lambda$1238,1240, 
Si {\small IV} $\lambda$$\lambda$1393,1402, 
and C {\small IV}$\lambda$$\lambda$1548,1551 doublets, though the 
N {\small V}$\lambda$$\lambda$1238,1240 feature is never detected at high S/N. 
The MagE spectrum is much higher spectral resolution than the GMOS spectrum, 
and would be the ideal dataset for comparison against synthetic spectra. However, 
we were limited to collecting 1 hr of integration time with MagE, and that taken at 
relatively high airmass and in highly variable seeing. As a result, the MagE data do 
not provide a superior comparison against the \texttt{S99} model spectra than the 
GMOS data, and so we focus on \texttt{S99} model-GMOS comparisons from here 
on out.

We generate an array of \texttt{S99} models assuming both continuous and 
instantaneous star formation scenarios. The continuous star formation models 
also spanning a range in metallicities from from solar to 2\% solar, and a range 
in ages of 2 Myr to 40 Myr. Our instantaneous models all assume a metallicity 
of 0.4Z$_{\odot}$ and span the range in 
ages (2--100 Myr), and also allow for variation in the maximum stellar mass 
formed, ranging from 30--120 M$_{\odot}$. In Figure~\ref{fig:pcygni} we plot 
two of the continuous star formation model synthetic spectra on top of 
the GMOS data in the regions surrounding the three strong P Cygni features 
noted previously (N {\small V}, S {\small IV} and C {\small IV}). The specific 
models plotted have metallicities of 0.4Z$_{\odot}$ -- in line with the metallicity 
we measure from nebular line diagnostics --  and ages of 2 Myr and 40 Myr. 
The \texttt{S99} synthetic spectra assuming an instantaneous burst of star formation 
are shown in Figure~\ref{fig:pcygni2}, also plotted for two stellar population ages 
(2 Myr and 10 Myr). None of the models with either continuous or instantaneous 
star formation can reproduce the combination of narrow and deep absorption 
features, the shape of the absorption trailing off blueward of the C {\small IV} 
line center, or the narrow and strong emission from Si {\small IV} and the 
broader strong emission from C {\small IV}. The strength of the observed 
Si {\small IV} and C {\small IV} emission could imply that these features are 
at least partly nebular in origin, rather than resulting entirely from the winds 
and atmospheres of massive stars.

\subsection{He {\small II} Emission}
\label{sec:heii}

He~{\small II} $\lambda$1640 emission appears in the GMOS and MagE spectra 
of \arcname. This feature appears in composite spectra of z $\sim$ 3 galaxies 
\citep{shapley2003} and in the spectrum of individual strongly lensed z $\sim$ 2-4  
galaxies \citep[e.g.;][]{cabanac2008,dessauges2011}, but typically appears as a 
broad emission feature that is associated with the winds of W-R stars, and exhibits 
velocity widths of $\sim$1000 \kms. However, the He~{\small II} $\lambda$1640 detected 
from \arcname~has a width of $\sim$330 \kms~in the GMOS spectrum, which matches the 
resolution of that data. The feature is too low S/N in the MagE spectrum to inform a 
multiple component fit to the velocity profile, but a simple gaussian fit to the line prefers 
a FWHM that is consistent with the resolution of the MagE spectra ($\sim$75 \kms), 
implying that a significant fraction of the emission does indeed originate from 
a narrow distribution in velocity space, and therefore is likely resulting from nebular 
emission rather than W-R winds.

Explaining such strong, narrow emission from He {\small II} requires extreme 
ionization. The He~{\small II}$\lambda$1640 line is extremely bright in \arcname, 
with a total flux ratio of He~{\small II} $\lambda$1640/H$\beta$ $\sim$ 0.17 after 
applying the extinction correction, and likely originates in part from nebular 
emission.  \citet{erb2010} note the presence of significant 
nebular He{\small II} $\lambda$1640 emission in a bright field-selected LBG at 
z $=$ 2.3, and find that high ionization parameter values are required to explain the 
ratio that they measure for He~{\small II} $\lambda$1640/H$\beta$ of 0.3, and an 
equivalent width of 2.7\AA.

We can also compare the equivalent width, W$_{1640}$, of 
He~{\small II} $\lambda$1640 in \arcname~against predictions for the Wolf-Rayet 
wind He~{\small II} $\lambda$1640 emission strength in the \texttt{S99} models 
described above; we measure W$_{1640} =$ 1.5 $\pm$ 0.15 \AA. The \texttt{S99} 
models only generate values this high for an extremely short-lived stretch 
(t$_{age} \sim$ 5--7 Myr) and only in models with solar metallicity (in strong 
disagreement with all of the metallicity diagnostics measured in 
$\S$~\ref{sec:oabundance}). And as noted previously, He~{\small II} $\lambda$1640 
emission originating from W-R winds would be much broader than the unresolved 
line that we see in the GMOS spectrum. Lastly, we also see no evidence of a P Cygni 
blue shifted absorption feature in the He~{\small II} $\lambda$1640 feature, which 
further argues against this line as originating entirely from W-R winds.

\section{Discussion}
\label{sec:discussion}

\subsection{Morphology of \arcname}

{\it HST} imaging enables us to examine the morphology of \arcname~via both the 
internal structure of the arc itself, as well as the much less distorted counter image. 
The counter image appears extremely irregular, and contains several distinct 
emission knots. The giant arc also includes multiple images of three distinct knots 
of emission, one of which is considerably redder than the rest of the arc. The central 
emission knot in the counter image also appears notably redder. While the knots 
have different IR-optical colors, all are extremely bright in the optical (rest-UV). This 
suggests that there is active star formation throughout \arcname, but that the redder 
central knot likely has a a larger underlying population of older stars. It is possible 
that the redder knot corresponds to the core of the galaxy, which possibly hosts the 
early stages of an assembling bulge, and that the bluer emission from the outskirts 
of the galaxy is preferentially tracing small but intense star forming regions.


\subsection{Comparison Against Low Redshift Galaxies}

\subsubsection{Strength of the C {\small III}] Feature}

Strong C {\small III}]$\lambda$$\lambda$1907,1909 emission is not ubiquitous in 
star forming galaxies at moderate redshift \citep[e.g., z $\sim$ 2 
composite;][]{shapley2003}. It is therefore interesting to compare \arcname\ 
against individual low redshift star forming galaxies with good UV spectra to 
try and understand the astrophysical conditions that are conducive to producing 
strong C {\small III}]$\lambda$$\lambda$1907,1909. The galaxy sample compiled by 
\citet{leitherer2011} (hereafter L11) is an excellent comparison set, with {\it HST} 
UV spectra of 46 star forming regions within 28 galaxies with z $<$ 0.06. 

\begin{figure}
\centering
\includegraphics[scale=0.443]{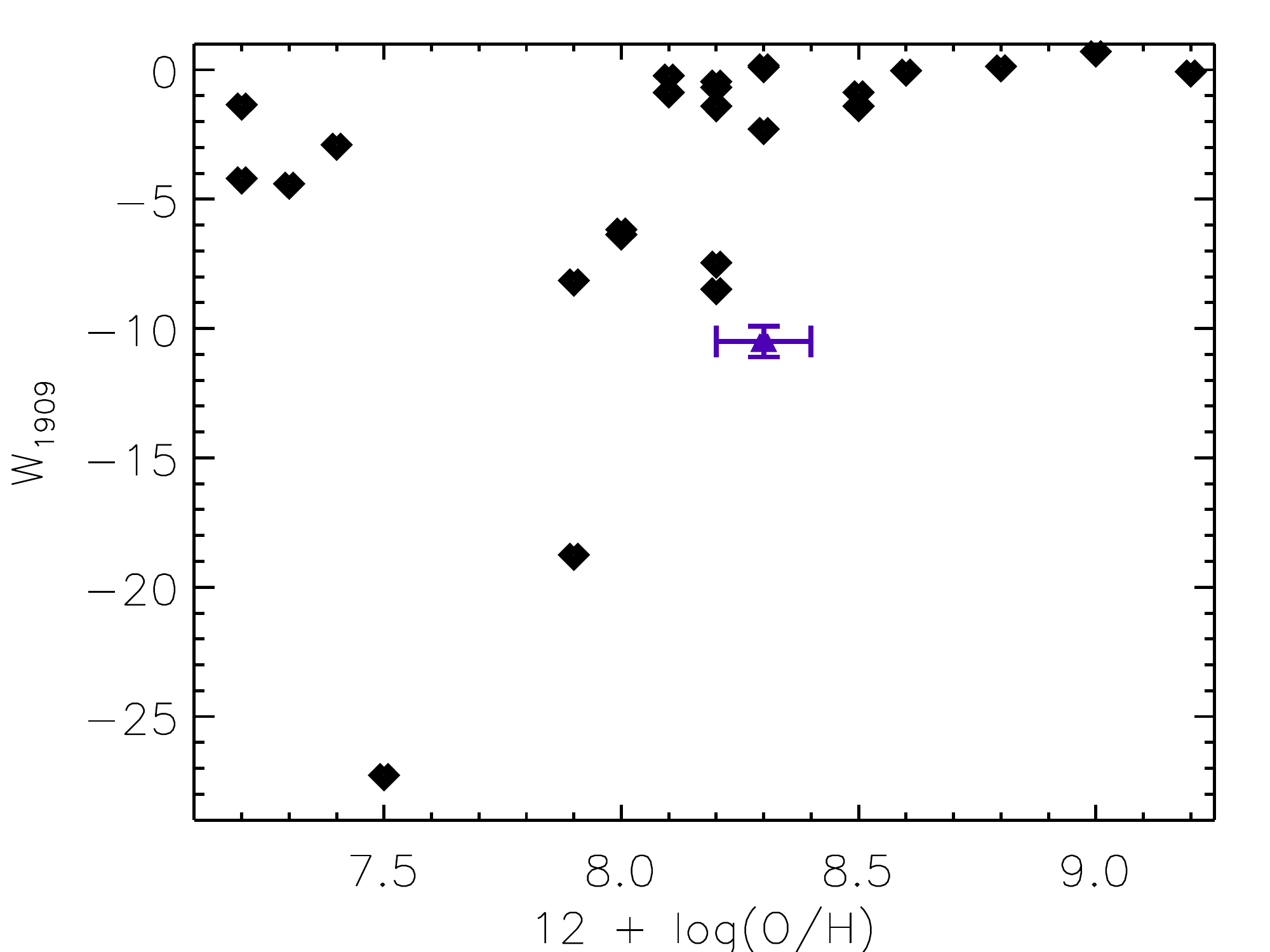}
\caption{\scriptsize{
Black notched diamonds show the equivalent width of the 
C {\small III}]$\lambda$$\lambda$1907,1909 doublet and metallicity 
for 25 galaxies in \citet{leitherer2011} where the line is included in the 
spectral coverage of the available data and there is a metallicity 
measurement available. The same values for \arcname~are indicated 
by the solid purple triangle.}}
\label{leitherer_ciii}
\end{figure}

We examine the individual spectra that were analyzed in L11, and measure the 
equivalent width, W$_{1909}$, of the \ciii$\lambda$$\lambda$1907,1909 line in 
the L11 spectra. In Figure~\ref{leitherer_ciii} we show the relationship between 
W$_{1909}$ and metallicity in the L11 sample; there is a dearth of galaxies with 
both a large W$_{1909}$ and high metallicity, indicating that 
\ciii$\lambda$$\lambda$1907,1909 emission may be suppressed in high 
metallicity star forming galaxies. Interestingly, \arcname~has a larger W$_{1909}$ 
than all but 2 of the 25 L11 galaxies, yet it has a metallicity, \logoh~$=$ 8.3, 
which coincides almost exactly with the apparent turn-off of strong \ciii~emission 
in the L11 sample. The \ciii~lines are forbidden and semi-forbidden 
transitions, so any such suppression must be the result of one or more indirect 
mechanisms, unlike the suppression of resonant lines such as Lyman-$\alpha$. 

A similar relationship has also recently been observed in low-mass high redshift 
galaxies \citep[Private communication;][]{stark2014prep}, where W$_{1909}$ 
appears to be correlated with the strength of Lyman-$\alpha$ emission.
In this context \arcname~is puzzling, in that is exhibits a strong DLA feature, which 
is opposite of what one should expect given a correlation between W$_{1909}$ 
and W$_{Ly-\alpha}$. The strength of \ciii~emission in \arcname~is difficult to 
understand in light of this galaxy's other observable properties and the apparent 
tendency for strong \ciii~emission to coincide with strong Ly-$\alpha$ emission and 
low metallicity. It would seem that the physics which dictate strong \ciii~cannot be 
summarized according to a simple correlation against a singe fundamental 
physical quantity (e.g., metallicity).


\subsubsection{Relative C/O, N/O, Si/O Enrichment}

With measurements of the relative abundance of C/O, N/O and Si/O we can compare 
the enrichment of the ISM in \arcname~against well-studied star forming galaxies 
at low redshift. \citet{garnett1995b} examine the relative C/O, C/N and N/O abundances 
of irregular/H{\small II} galaxies at low redshift. \arcname~fits nicely onto the sequence 
of C/O vs. \logoh~ and C/N vs. \logoh~that \citet{garnett1995b} observe.

The relative abundances of Si and O were also studied by \citet{garnett1995a}. Si and 
O should be produced in the same stars, and therefore their ratio is not expected to 
vary strongly with metallicity. Variations can occur, however, if Si depletes onto 
dust grains in the ISM, rendering it unobservable via observation of line emission 
from ionized H{\small II} regions. Our measurement of the Si/O abundance in \arcname~ 
is somewhat lower than the stable value observed by \citet{garnett1995a}, but the large 
uncertainties prevent us from making a strong statement about whether or not Si 
depletion by the formation of silicate dust grains may truly be taking place in \arcname.

Our abundance measurements indicate that \arcname\ has elemental enrichment 
properties that are generally in-line with observations of irregular star 
forming galaxies at z $\sim$ 0.


\subsection{P Cygni Features}
\label{sec:pcygni}

Based on the strength of the P Cygni features in \arcname, we expect the extremely 
low metallicity models to do a poor job of reproducing the observed features, and this 
holds true. Models with metallicities of 0.4Z$_{\odot}$ result in the best agreement 
with our data, which is encouraging given that we measure the metallicity 
to be $\sim$0.4Z$_{\odot}$ from nebular emission line diagnostics. However, none 
of the synthetic spectra that we generate using \texttt{S99} produce the combination 
of P Cygni features that we observe in of \arcname. The most challenging features 
to explain are the narrow shape and exceptional strength of the blueshifted 
absorption features, as well as the strength of the redshifted C {\small IV} emission 
and the odd, strong emission from S {\small IV}$\lambda$1393. Qualitatively these 
features indicate the presence of an extremely young population of massive 
stars. Some recent work has also explored the ways in which stellar rotation 
can affect the spectra of massive stars, and find that including rotation effects in the 
modeling of stellar spectra can result in a increase in the amount of ionizing 
radiation, hardening of the ionizing radiation, and stronger profiles for some 
lines, including the UV Si {\small IV} and C {\small IV} P Cygni features 
\citep{levesque2012,leitherer2014}.


Our results here are reminiscent of the difficulty that other studies have encountered 
comparing synthetic spectra against observations of galaxies at z $\gtrsim$ 2 
\citep{pettini2000,shapley2003,quider2009,erb2010}. As suggested by \citet{erb2010}, 
the excess emission features could be explained, at least in part, by 
nebular emission. The presence of nebular emission from these transitions does, 
however, agree with the properties of the He {\small II}$\lambda$1640 emission 
line discussed above, and would imply an extremely hard ionizing radiation field 
from massive O stars. 


One possible explanation for what we observe in \arcname~is that the integrated 
spectrum that we measure across a wide range of wavelengths is a blend of the 
properties from several different star forming regions embedded within 
the galaxy. Different star forming regions could be generating extreme spectral 
features that dominate the observable signal in some regions of the spectrum 
(e.g., the rest-UV) while only contributing in part to other parts of the spectrum 
(e.g., the rest optical). The superposition of a ``normal'' star forming galaxy with 
small, highly magnified regions of intense and short-lived star formation could be 
responsible for generating the complex spectral features that defy reproduction 
by a simple monolithic synthetic population of stars. There is, in fact, a body of 
published work that uses real \citep{kobulnicky1999,james2013a,james2013b} 
and simulated \citep{pilyugin2012} observations of well-studied low-redshift 
star forming galaxies to explore the biases that can result from simple analyses of 
the integrated spectra of distant galaxies. Recent observations of different line of 
sight within a single lensed star forming galaxy also shows clear evidence of 
different physical conditions in different star forming regions \citep{rigby2014}. 
There is clear evidence to suggest that spatially resolved spectroscopy will be 
essential for constraining some of the physical properties of high redshift 
star forming galaxies.

\subsection{Interpreting High Ionization Features in the Rest-UV}

It is physically unlikely that all of our measured He~{\small II}$\lambda$1640 line flux 
is nebular in origin, as that would require ionization parameters $> -1$ according to 
the models explored by \citet{erb2010}. Even assigning only $\sim$25\% of the 
flux to nebular emission requires an ionization parameter, log(U) $>$ $-$2 for 
metallicities in the range that we measure for \arcname, according to the models 
explored by \citet{erb2010}. A physical picture in which there are one or more 
extreme star forming region(s) within \arcname~is also consistent with evidence 
that unresolved starbursts have a maximum ionization parameter, log(U) $\sim$ 
$-$2.3 \citep{sherry2012}, implying that diagnostics indicating significantly larger 
values of log(U) are likely to be sampling individual extreme star forming regions 
within galaxies \citep[e.g.,][]{snijders2007,indebetouw2009}. 

Another possible explanation for the strong He~{\small II}$\lambda$1640 line is 
the presence of an exceptionally hard ionizing spectrum. The ionization parameter, 
$U$, only describes the intensity of ionizing radiation; a very hard ionizing spectrum 
with relatively low intensity can generate a large amount of nebular He~{\small II} 
emission even given a low ionization parameter. An initial mass function (IMF) that 
generates more of the hottest, most massive stars, for example, could account for a 
harder ionizing spectrum than considered by the models of \citet{erb2010}. 
However, arguments that are based on the vague, qualitative implications of varying 
the IMF are distasteful and presumptive when one considers that the spectral properties 
of the most massive stars are not well understood, especially at low metallicity 
\citep[e.g.,][]{rigby2004}.

The strong, narrow, He~{\small II} $\lambda$1640 emission 
that we detect in \arcname~presents a puzzle. It either casts serious doubt on 
the ionization parameter diagnostics that we compute and rely on throughout 
$\S$~\ref{sec:ionabundance}, or it suggests a hard ionizing spectrum that is 
difficult to reconcile with the electron temperature constraints available from 
rest-frame optical emission lines. 

Returning to the explanation that we put forth in the previous subsection, it 
seems likely that the integrated spectrum of \arcname~is too complex to be 
described by a simple stellar population, or a single list of parameters (i.e., 
a single ionization parameter, electron temperature, metallicity, etc). Rather, 
it is possible that the [O {\small III}]$\lambda$$\lambda$4960,5008, 
[O {\small II}]$\lambda$$\lambda$3727,3729, [Ne {\small III}]$\lambda$3869, and 
He~{\small II} $\lambda$1640 lines are in fact blends of multiple components which 
are originating from different physical regions within the galaxy. A relatively small 
but extreme region of recent star formation with a large population of hot massive 
stars could, for example, produce strong He~{\small II} emission, while other more 
``normal'' regions within the galaxy could be responsible for weighting the 
measured [O {\small III}]/[O {\small II}] and [Ne {\small III}]/[O {\small II}] line ratios 
toward more typical values. 


In cases where we are studying the physical properties of strongly lensed star 
forming galaxies, it is easy to imagine that differential magnification effects 
across the surface of the background galaxy can generate integrated observable 
quantities that are weighted toward the highest surface brightness regions within 
the lensed galaxy, which would naturally tend to be the regions of most intense 
unobscured star formation \citep[e.g.,][]{er2013}. This issue is not, of course, unique 
to strongly leaned galaxies at high redshift. Studies of the integrated properties of 
any distant galaxy must, ultimately, account for the fact that properties such as 
metallicity, SFR and ionization are not held uniform throughout a given galaxy. The 
fundamental scale of star formation in the universe is much smaller than a galaxy, 
and individual galaxies can (and should) therefore host a range of star forming 
regions with physically different properties. This point is emphasized by the few 
published studies of lensed galaxies with IFU units in the NIR 
\citep[e.g.,][]{stark2008,wuyts2014}. Spatially resolved/IFU spectroscopy of distant 
sources is a powerful tool for confronting some of the dangers that result from 
analysis (or over-analysis) of integrated spectra of these galaxies. It is also 
sensible that the diversity of properties within an individual galaxy could vary more 
dramatically in the earlier universe, when star formation was still ramping up to its 
peak and galaxies had yet to assemble a majority of their stars relative to 
galaxies in the present epoch.

\section{Summary and Conclusions}
\label{sec:conclusions}

We present a detailed analysis of optical and NIR imaging and spectroscopy of an 
exceptionally bright strongly lensed galaxy at z $=$ 3.6252. \arcname~is among the 
best-characterized star forming galaxies at z $>$ 2, and the highest redshift galaxy 
with its properties measure from high S/N rest-frame UV \emph{and} optical spectra. 
The observations and analyses presented here are a step toward improving 
observational constraints on the internal astrophysics in high redshift,  vigorously 
star-forming galaxies. Our primary results are:

\begin{enumerate}

	\item[--] \arcname~is a moderate-metallicity (Z $=$ 0.4Z$_{\odot}$) 
	moderately low-mass (log(M$_{*}$/M$_{\odot}$) $=$ 9.5) galaxy, with 
	star  formation rates of 55 $\pm$ 25 and 84 $\pm$ 24 M$_{\odot}$ yr$^{-1}$ 
	measured from nebular [O {\small II}]$\lambda$$\lambda$3727 and H-$\beta$ 
	emission, respectively, implying that vigorous 
	starbursts are taking place in one or more regions within the galaxy. 
		
	\item[--] Several derived physical characteristics of \arcname, including estimates of 
	relative elemental abundances and the strong \ciii$\lambda$$\lambda$1907,1909 
	emission are in-line with well-studied star forming galaxies at z $\sim$ 0. Other 
	features such as the strong damped Lyman-alpha absorption, however, are 
	surprising given the strong C {\small III} emission.

	\item[--] Some features in the UV spectrum -- the P Cygni lines and 
	strong He~{\small II} emission -- indicate a strong ionizing field and/or a very 
	high ionization parameter in conflicts with rest-frame optical diagnostics. 
	Attributing strong He~{\small II} emission in the UV to a high ionization parameter, 
	requires log(U) $>$ $-$2, whereas the rest-frame optical nebular lines prefer a more 
	``normal'' value of log(U) $< -2.05$.

	\item[--] Our work here is a reminder that as the quality of data improve for high 
	redshifts galaxies, it is essential that the subsequent analyses are aware of and 
	account for the systematic effects that result from measuring properties from spectral 
	features that are a combination of emission originating from different regions within 
	galaxies. The fundamental mode of star formation in the universe operates on scales 
	much smaller than individual galaxies. Studies of strongly lensed galaxies, therefore, 
	provide a unique opportunity to probe the spatial variance of star formation in 
	galaxies during the era of peak star formation.
		
\end{enumerate}

~~~~~

\arcname~is a prime target for more extensive follow-up observations. High 
resolution optical spectra, in particular, would help to answer many of the 
outstanding questions about the nature of the star formation, ISM, and the 
properties of the population of massive and W-R stars within this galaxy.

\acknowledgments{We thank Sally Heap and Dan Stark for interesting 
and helpful discussions that improved this paper, as well as the anonymous 
referee who provided very thoughtful and constructive feedback. This work 
was supported by the National Science Foundation through Grant AST-1009012, 
by NASA through grant HST-GO-13003.01 from the Space Telescope Science 
Institute, which is operated by the Association of Universities for Research in 
Astronomy, Incorporated, under NASA contract NAS5-26555, and also by the 
FIRST program ``Subaru Measurements 
of Images and Redshifts (SuMIRe)'', World Premier International Research 
Center Initiative (WPI Initiative), MEXT, Japan, and Grant-in-Aid for Scientific 
Research from the JSPS (23740161).}

\bibliographystyle{apj}
\bibliography{/Users/mbayliss/astro/master_bibliography}

\end{document}